\documentclass[twocolumn,pre,superscriptaddress,aps,nofootinbib]{revtex4}

\usepackage{url,psfrag,graphicx}
\usepackage{dcolumn}
\usepackage{amsmath,amssymb}
\usepackage{bm}
\usepackage{hyperref}
\usepackage{float,epsfig,color}
\usepackage{times}
\usepackage{array}
\usepackage{braket}
\usepackage{diagbox}
\usepackage{lscape}
\usepackage{multirow}
\usepackage{stackrel}
\usepackage{verbatim}
\usepackage{wasysym} 
\usepackage{xcolor}
\usepackage{soul}

\usepackage{hyperref}

\DeclareMathAlphabet      {\mathitbf}{OML}{cmm}{b}{it}

\begin{document}

\title{Spreading fronts of wetting liquid droplets: microscopic simulations and universal fluctuations}

\author{J. M. Marcos}
\affiliation{Departamento de F\'{\i}sica, Universidad de Extremadura, 06006 Badajoz, Spain}
\address{Instituto de Computaci\'on Cient\'{\i}fica Avanzada de Extremadura (ICCAEx), Universidad de Extremadura, 06006 Badajoz, Spain}
\author{P. Rodr\'iguez-L\'opez}
\address{\'Area de Electromagnetismo and Grupo Interdisciplinar de Sistemas Complejos (GISC), Universidad Rey Juan Carlos, 28933 M\'ostoles, Spain}
\author{J. J. Mel\'endez}
\affiliation{Departamento de F\'{\i}sica, Universidad de Extremadura, 06006 Badajoz, Spain}
\address{Instituto de Computaci\'on Cient\'{\i}fica Avanzada de Extremadura (ICCAEx), Universidad de Extremadura, 06006 Badajoz, Spain}
\author{R. Cuerno}
\affiliation{Departamento de Matem\'aticas and GISC, Universidad Carlos III de Madrid, 28911 Legan\'es, Spain}
\author{J. J. Ruiz-Lorenzo}
\affiliation{Departamento de F\'{\i}sica, Universidad de Extremadura, 06006 Badajoz, Spain}
\address{Instituto de Computaci\'on Cient\'{\i}fica Avanzada de Extremadura (ICCAEx), Universidad de Extremadura, 06006 Badajoz, Spain}

\date{\today}

\begin{abstract}
We have used kinetic Monte Carlo (kMC) simulations of a lattice gas to study front fluctuations in the spreading of a non-volatile liquid droplet onto a solid substrate. Our results are consistent with a diffusive growth law for the radius of the precursor layer, $R \sim t^{\delta}$, with $\delta \approx 1/2$ in all the conditions considered for temperature and substrate wettability, in good agreement with previous studies. The fluctuations of the front exhibit kinetic roughening properties with exponent values which depend on temperature $T$, but become $T$-independent for sufficiently high $T$. Moreover, strong evidences of intrinsic anomalous scaling have been found, characterized by different values of the roughness exponent at short and large length scales. Although such a behavior differs from the scaling properties of the one-dimensional Kardar-Parisi-Zhang (KPZ) universality class, the front covariance and the probability distribution function of front fluctuations found in our kMC simulations do display KPZ behavior, agreeing with simulations of a continuum height equation proposed in this context. However, this equation does not feature intrinsic anomalous scaling, at variance with the discrete model.
\end{abstract}

\maketitle

\section{Introduction}\label{sec:intro}
The expansion of non-volatile liquid droplets deposited onto flat surfaces (to which we will refer hereafter simply as spreading) appears in many processes of scientific or technological relevance \cite{Reiter2018}. This fact justifies the intensive research that fluid spreading has originated during the last decades. Being critically conditioned by the wetting properties of the chosen fluid/substrate combination which is considered in each case, the dynamics of spreading is a complex phenomenon that has been studied from a number of approaches \cite{Starov2007}. At the {\em macroscopic} scale, the radius of a droplet, $R_{\rm d}$, depends on time $t$ as a power law, $R_{\rm d} \sim t^{\delta_{\rm d}}$, where $\delta_{\rm d}$ depends on the physical-chemical characteristics of the droplet and the substrate, and ranges between $\delta_{\rm d} = 1/10$ (a result known as Tanner's law \cite{Tanner1979}) and $\delta_{\rm d} = 1/7$ \cite{Bonn2009}. In general, the macroscopic behavior of the system is accurately described by hydrodynamics, except in the complete wetting regime. For this case, experimental results evince the occurrence of a so-called precursor layer, with a {\em microscopic} thickness of about one or a few molecules across, which precedes the macroscopic droplet and spreads out ahead of it \cite{Popescu2012}. The radius of the precursor layer, $R(t)$, grows as $R \sim t^{\delta}$ with $\delta \approx 1/2$ \cite{Heslot1989,Valignat1998}, thus featuring much faster growth than that of the droplet itself. Besides, there is a wide consensus about the universality of the behavior of the precursor layer, in the sense that its phenomenology does not depend much on the geometry of the experimental setup or on the composition of the wetting liquid \cite{Bonn2009}. Actually, precursor dynamics has been identified in systems of a very different nature, such as e.g.\ the expansion of cellular aggregates \cite{Douezan2011}. On the other hand, the spreading behavior of ultrathin liquid precursor films is recently being shown to enable experimental control at very small distances, as demonstrated e.g.\ for precise positioning of sub-10 nm particles into lithographically defined templates \cite{Asbahi2015}.

The occurrence of the precursor layer at molecular scales poses limitations to the theoretical description of the process by hydrodynamics, and has motivated the development of more detailed (and sophisticated) microscopic models, see e.g.\ Ref.\ \cite{Popescu2012} for a review. Depending on the type of property which needs to be addressed and the degree of quantitative precision which is required, these range from molecular dynamics \cite{DeConinck2008,Luo2020} to Monte Carlo (MC) simulations \cite{Popescu2012}. The latter are particularly well suited to the study the effect of fluctuations in the large-scale (long time and large distance) behavior of the system, which is the focus of our present work.

Following the pioneering work by de Gennes and Cazabat \cite{deGennes1990}, most MC studies of fluid spreading model the short-ranged interaction between the substrate and the liquid by a van der Waals-like term which decreases in absolute value with the vertical distance $Z$ from the substrate, typically in the form $-A Z^{-3}$, where $A$ is the Hamaker constant. \footnote{To avoid confusion with standard notation for the so-called dynamic exponent $z$ to be introduced below, we capitalize the vertical coordinate $Z$ in 3D space.} Here, we will focus on a lattice gas model developed originally by Lukkarinen \textit{et al.} \cite{Lukkarinen1995}: in essence, it is a modification of the Ising model with an external field, with the occupation numbers of the lattice sites, $n(\bm{r})$, playing the role of the spins. The modified Hamiltonian is written as
\begin{equation}
    \mathcal{H}= -J \sum_{\langle \bm{r}, \bm{s} \rangle} n(\bm{r},t)n(\bm{s},t) - A \sum_{\bm{r}}\frac{n(\bm{r},t)}{Z^3},
    \label{eq:energy}
\end{equation}
where the first term describes the interactions between the liquid particles and their nearest neighbors, the second one accounts for the interaction with the substrate, characterized by a Hamaker constant $A > 0$, and $\mathbf{r}=(x,y,Z)$ denotes position in a three-dimensional (3D) square lattice. This model is consistent with the $R \sim t^{1/2}$ law for the precursor layer, as demonstrated by kinetic Monte Carlo (kMC) simulations by Abraham \textit{et al.} \cite{Abraham2002}. These simulations also showed that the spatiotemporal fluctuations of the front position display so-called kinetic roughening behavior \cite{Barabasi1995,Krug1997} and identified the associated critical exponent values in the low-temperature regime. Moreover, agreement was assessed with predictions from a nonlinear continuum equation for the front position which was derived from a stochastic moving boundary formulation of the problem, in which the celebrated Kardar-Parisi-Zhang (KPZ) nonlinearity \cite{Kardar1986,Barabasi1995,Krug1997} played a conspicuous role. Note that several theoretical and experimental breakthroughs, precisely in the context of the KPZ universality class of surface kinetic roughening, have boosted during the last decade our understanding of this type of critical behavior far from equilibrium, see e.g.\ \cite{HalpinHealy2015,Takeuchi2018} for reviews. Thus, nowadays universal behavior is known in this context to reach beyond the values of the critical exponents, extending to universal forms of the probability density function (PDF) (one-point statistics) and covariance (two-point statistics) of front fluctuations. In the case of the one-dimensional (1D) KPZ universality class, these are related with the Tracy-Widom family of PDFs for extremal eigenvalues of random matrices \cite{Kriecherbauer2010,Corwin2012} and with the covariance of the so-called Airy processes \cite{HalpinHealy2015,Takeuchi2018}, respectively. Moreover, such a strong form of universality extends to classes other than 1D KPZ, both linear and nonlinear \cite{Carrasco2016,Carrasco2019}. And remarkably, it applies to the spatiotemporal fluctuations of a host of low-dimensional non-equilibrium systems with strong correlations, which are not necessarily interfaces, from active matter \cite{Chen2016} to quantum matter \cite{He2017} and from quantum dots to human cells \cite{Makey2020}.

With respect to the results of Ref. \cite{Abraham2002}, Harel and Taitelbaum \cite{Harel2018,Harel2021} have very recently performed additional extensive kMC simulations of the modified Ising model, Eq.\ \eqref{eq:energy}, to study the influence of the Hamaker constant and temperature ($T$) on the dynamics of wetting. The results of these studies differ noticeably from those in \cite{Abraham2002}. Now the $R \sim t^{1/2}$ growth law is not obtained, but rather a relation of the form $R \sim t^{\delta}$, with $\delta$ depending strongly with $A$ and $T$ and far from the diffusive 1/2 value. Such a lack of universality affects the front fluctuations as well, whose scaling exponents also depend sensitively on the system parameters in the simulations of \cite{Harel2018,Harel2021}. These results are unexpected, as they conflict with independent simulations of the same model published by different authors \cite{Lukkarinen1995,Abraham2002,Luo2019}. Moreover, the conclusion in \cite{Harel2018,Harel2021} that $0.13 \lesssim \delta \lesssim 0.28$ for the various parameters studied directly challenges the very suitability of the modified Ising model to describe the spreading process, as these values are well off the experimentally measured $\delta\approx 1/2$ value \cite{Bonn2009}.

In view of the recent progress in the characterization of kinetic roughening universality classes ---which is susceptible of experimental assessment, as exemplified by the KPZ universality class of 1D \cite{Takeuchi2012} and 2D \cite{Almeida2014,Halpin-Healy2014,Orrillo2017} interfaces---, we believe that a thorough study is needed for the spreading of thin films, in which the $R \sim t^{\delta}$ law is correlated with and complemented by the type of front fluctuations that can be expected as a function of system parameters. To this end,
\begin{itemize}
\item We revisit the lattice gas model, Eq.\ \eqref{eq:energy}, by performing new extensive kMC simulations for wide ranges of values for the Hamaker constant $A$ and temperature $T$. We will compute the front position and its fluctuations in time and space both, for the precursor layer, and for the next layer on top of it, usually called supernatant \cite{Abraham2002}.

\item From these variables we will investigate the dynamical evolution of the layers and its dependence on the values of $A$ and $T$, addressing the occurrence of universal behavior in the light of recent developments for the KPZ universality class, and also of the results reported in \cite{Harel2018,Harel2021}.

\item While universal behavior does take place, its full understanding will require introducing certain refinements over its simplest manifestations, in particular the so-called anomalous kinetic roughening \cite{Krug1997,Lopez1997} behavior.

\item Some unexpected behaviors, like the occurrence of KPZ one- and two-point statistics together with {\em non}-KPZ exponent values, will be contrasted with similar behavior obtained from continuum descriptions of spreading dynamics \cite{Abraham2002}.

\item Our results on kinetic roughening behavior will be brought into context in the light of analogous behavior found for other systems displaying kinetic roughening.

\end{itemize}

To address all these points we have organized the paper as follows. Section II recalls basic details on simulation procedures for the discrete model, Eq.\ \eqref{eq:energy}, and provides the definitions of the quantities that will be measured. Our numerical results are reported and discussed in Sec.\ III, which is finally followed by a summary and our conclusions in Sec.\ IV. Details on the parameter values considered in our simulations, our statistical data analysis, and some of our numerical results on exponents values are provided in an appendix.

\section{Simulation details and definitions} \label{observables}
The microscopic driven Ising lattice gas model considered here consists of two overlapping $2D$ rectangular layers, of dimensions $L_{x} \times L_{y}$. Each node of the square lattice $\mathitbf{r}= (x, y, Z)$ can be occupied by at most one particle at any time, so the occupation number $n(\mathitbf{r},t)$ may take the values 0 or 1. The lower $(Z = 1)$ and upper $(Z=2)$ layers are called precursor and supernatant, respectively, and the substrate on which the droplet expands is located at $Z = 0$. Periodic boundary conditions are employed in the $y$-direction, as in Ref.\ \cite{Abraham2002,Harel2018,Harel2021}. Note, the choice of boundary conditions is known not to influence universal properties, such as the values of exponents like $\delta$ \cite{Popescu2012}, or those characterizing the kinetic roughening behavior \cite{Barabasi1995,Krug1997} that may occur in the system.

The first $(x=0)$ column of both layers represents a fluid reservoir (the macroscopic droplet). Initially only these cells are occupied. If, due to an exchange, any of the cells of the reservoir becomes empty, it is instantaneously refilled. On the other hand, if at any later stage a particle occupies the last column of the lattice, the particle is assumed to escape from the system.

\begin{figure}[!t]
\centering
\includegraphics[width=0.45\textwidth]{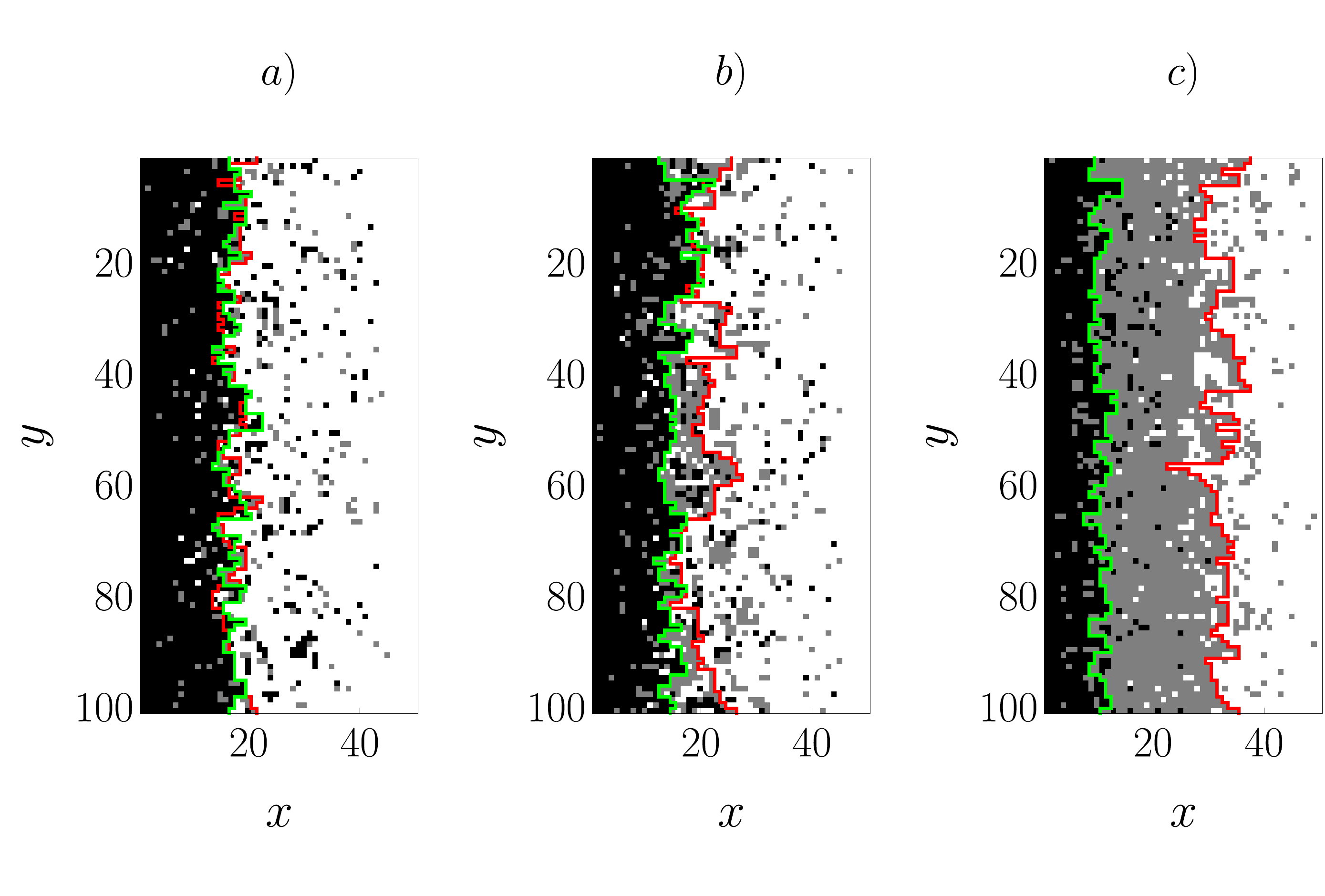}
\caption{(Color online) Top views of three snapshots of the lattice gas model for increasing values of the Hamaker constant $A$, left to right. Occupied cells in the precursor and supernatant layers are in gray and black, respectively, with the red and green lines delimiting the corresponding fronts; empty cells are uncolored. Parameters used are $J = 1$, $T = 1$, $L_{y} = 100$, $L_{x} = 50$, and a) $A = 0.1$, b) $A = 1$, and c) $A = 10$. The three snapshots were taken at the same simulation time. All units are arbitrary.}
\label{fig:grafCombT}
\end{figure}

\begin{figure}[!t]
\centering
\includegraphics[width=0.45\textwidth]{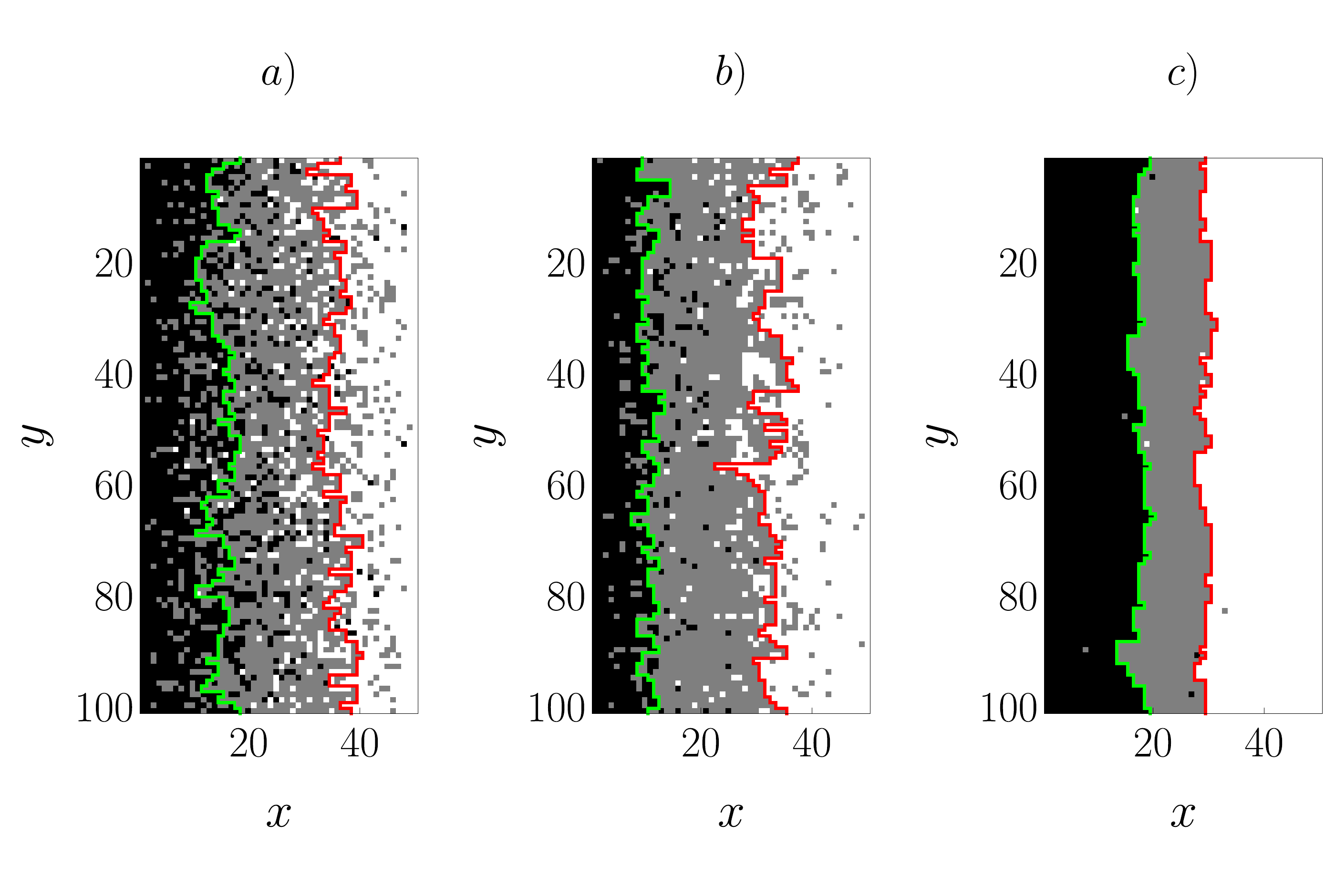}
\caption{(Color online) The same as in Fig.\ \ref{fig:grafCombT} but for for decreasing values of the temperature $T$, left to right. Specifically, $J = 1$, $A=10$, $L_{y} = 100$, $L_{x} = 50$, and a) $T = 3$, b) $T = 1$, and c) $T = 1/3$. All units are arbitrary.}
\label{fig:grafCombA}
\end{figure}

The total energy of the system is given by Eq.\ \eqref{eq:energy}, defined in terms of $A$ and $J$. From the physical point of view, the most interesting values for the pairs $(A,J)$ are those for which $J/k_{B}T$ is large enough to achieve a high degree of involatility and $A/k_{B}T$ is large enough to be in the complete wetting regime \cite{Abraham2002}. From Eq.\ \eqref{eq:energy}, it is clear that the lowest energy is achieved for the smallest value of $Z$, which indicates that the occupation of the precursor layer is energetically favorable. This preferential occupation is enhanced for $A \gg J$, in which case one would expect the bottom layer to grow faster than the upper one. On the contrary, when $J$ is dominant, one could expect that both layers grow at the same speed. From now on, we choose physical units such that $k_B=1$ and remain arbitrary otherwise. Figure \ref{fig:grafCombT} shows top-view snapshots of the system obtained for fixed $J$ (or, alternatively, $T$) and several values of the Hamaker constant. Note how the front roughness, defined in detail below, increases with the temperature of the system, for a given value of $A$ (Fig.\ \ref{fig:grafCombA}).

The evolution of the system has been simulated by continuous-time Monte Carlo Kawasaki local dynamics \cite{Newman1999}, which is described in Appendix~\ref{details}. At each time, a particle belongs to the precursor (or the supernatant) film if there are nearest-neighbor connections filled with particles all the way back to the droplet reservoir. Thus, for fixed $y$, the size of each layer, or front, $h(y,t,Z)$ is defined as the highest value of $x$ at which a cell is occupied. The average distance of the interface to the fluid reservoir (average front position or height) is then
\begin{equation}
    \bar{h}(t,Z)=\frac{1}{L_{y}}\sum_{y}h(y,t,Z)\,.
    \label{eq:distanciaProm}
\end{equation}
On the other hand, the front width (or roughness) at each layer, $w(L_y,t,Z)$, is defined as the standard deviation of the front \cite{Barabasi1995,Krug1997},
\begin{equation}
	\label{eq:width}
	w^2(L_y,t,Z)=\left\langle \overline{[h(y,t,Z)-\bar{h}(t,Z)]^2} \right\rangle,
\end{equation}
where we have used the notation $\overline{O(t,Z)} \equiv (1/L_y)\sum_y O(y,t,Z)$ for the average of a given observable $O(y,t,Z)$ defined at the position of the front on each layer. Furthermore, we denote by $\langle (\cdots) \rangle$ the average over different realizations of the system. Hereafter, we will denote this observable as $w^2(L_y,t)$ or simply by $w^2(t)$. Finally, we define the skewness $S$ and the kurtosis $K$ as functions of the local height fluctuation $\delta h(y,t)=h(y,t)-\bar{h}(t)$, namely, 
\begin{equation}\label{Eq:Skewness}
S~=~ \frac{\langle \delta h(y,t)^3 \rangle_c}{\langle \delta h(y,t)^2 \rangle_c^{3/2}}
\end{equation}
and
\begin{equation}\label{Eq:Kurtosis}
K=\frac{\langle \delta h(y,t)^4 \rangle_c}{\langle \delta h(y,t)^2 \rangle_c^{2}},    
\end{equation}
where $\langle (\cdots) \rangle_c$ denotes the cumulant average. 

In what follows, we will omit the layer $Z$ index, whose value will be clear from the context.

To describe the dynamical evolution of the front, two additional (equal-time) space correlation functions will be considered, namely, the height covariance
\begin{equation}
    	C_1(r, t) = \frac 1{L_y} \sum_y \langle h(r+y, t) h(y, t) \rangle-\langle \bar{h}(t)\rangle^2
    \label{eq:correlation_1}
\end{equation}
and the height-difference correlation function
\begin{eqnarray}\nonumber
	\label{eq:correlation_2}
    	C_2(r, t)&=&\frac 1{L_y} \sum_y \left\langle [h(y+r, t)-h(y, t)]^2 \right\rangle 	  \\
	    &=& 2\langle \bar{h}^2(t)\rangle-\frac 2{L_y}\sum_y \langle h(r+y, t)h(y, t) \rangle \,,
\end{eqnarray}
where the sum spans all $y$ values. While $C_1(r,t)$ will be used to assess universal fluctuation properties, $C_2(r,t)$ will allow us to evaluate the correlation length $\xi(t)$ along the front profile.

Figures \ref{fig:grafCombT} and \ref{fig:grafCombA} show examples of fronts in both layers. The average front (for $Z=1$, our estimate for the radius $R(t)$ of the precursor layer) is expected to grow as $ \bar{h}(t)\sim t^\delta$, with $\delta = 1/2$ for all parameter conditions \cite{Harel2018,Abraham2002}. On the other hand, under kinetic roughening conditions, the roughness $w(L_y,t)$ satisfies the so-called Family-Vicsek (FV) scaling law \cite{Barabasi1995,Krug1997,Abraham2002,Harel2018}
\begin{equation}
	\label{eq:w}
	w(L_y,t)=t^{\beta}f\left( t/L_y^z \right),
\end{equation}
where $w\sim t^{\beta}$ for $t \ll L_y^z$ and the saturation value of the roughness for larger times is $w_{\mathrm{sat}}\sim L_y^{\alpha}$ for $t \gg L_y^z$, so that $\alpha=\beta z$. Here, $\alpha$ is the roughness exponent, which is related with the fractal dimension of the front \cite{Barabasi1995} and characterizes the fluctuations of the front roughness at (large) scales comparable with the system size, as implied by the $w_{\mathrm{sat}}\sim L_y^{\alpha}$ scaling.
Further, $\beta$ in Eq.\ \eqref{eq:w} denotes the growth exponent and $z$ is the so-called dynamic exponent, which quantifies the power-law increase of the lateral correlation length along the front \cite{Barabasi1995,Tauber2014},
\begin{equation}
    \label{eq:correlation_length}
    \xi(t) \sim t^{1/z}\,.
\end{equation}
In the context of scaling behavior, a correlation function like $C_2(r,t)$, which is quadratic in the height, is expected to scale with lateral distance as
\begin{equation}
    \label{eq:correlation_length_loc}
    C_2(r,t) \sim r^{2\alpha_{\rm loc}}\,,
\end{equation}
for distances below the correlation length, i.e.\ such that $r\ll\xi(t)$. Here, $\alpha_{\rm loc}$ is a roughness exponent which is measured at local distances smaller than the system size. Under the FV scaling hypothesis, $\alpha_{\rm loc}=\alpha$ \cite{Barabasi1995,Krug1997}. However, there are more complex scaling scenarios, termed anomalous scaling \cite{Sarma1994,Plischke1993,Schroeder1993,Krug1997,Lopez1997,Ramasco2000}, for which $\alpha_{\rm loc}\neq\alpha$. As will be discussed below, our kMC simulations are consistent with so-called intrinsic anomalous scaling, wherein the height-difference correlation function behaves as \cite{Lopez1997}
\begin{equation}
    C_2(r,t)=r^{2\alpha} g(r/\xi(t)) \,,
    \label{eq:corr_length}
\end{equation}
where $g(u) \sim u^{-2(\alpha-\alpha_\mathrm{loc})}$ for $u\ll 1$ and $g(u) \sim u^{-2\alpha}$ for $u\gg 1$. For later use, we define $\alpha^\prime \equiv \alpha-\alpha_\mathrm{loc}$. Simple FV scaling corresponds to $\alpha'=0$, so that $g_{\rm FV}(u) \sim$ constant for $u\ll1$ \cite{Barabasi1995,Krug1997}. In this case the surface is a self-affine fractal and there is a single roughness exponent which characterizes both, small and large-scale fluctuations in space. In the presence of intrinsic anomalous scaling, the condition of self-affinity is not fulfilled, and the local and global space fluctuations do not scale with the same exponent. Note that in this case there are three independent exponents (rather than two, e.g.\ $\alpha$ and $z$, for FV scaling) characterizing the scaling behavior, e.g.\ $\alpha$, $\alpha_{\rm loc}$, and $z$ \cite{Lopez1997}. Anomalous scaling can be particularly well studied  by means of the surface structure factor $S(k,t)=\langle |\delta h_{k}(t)|^2 \rangle$ \cite{Siegert1996,Lopez1997}, where $\delta h_{k}(t)$ is the Fourier transform of $\delta h(y,t)$ and $k$ is 1D wave-vector. In the case of a 1D interface displaying intrinsic anomalous scaling, one has
\begin{equation}
    S(k,t)=|k|^{-(2\alpha+1)} s(|k|\xi(t)) \,,
    \label{eq:Sk}
\end{equation}
where $s(u) \sim u^{2\alpha+1}$ for $u\ll 1$ and $s(u) \sim u^{2\alpha'}$ for $u\gg 1$ \cite{Lopez1997}. This scaling behavior implies that, for long enough times, the structure factor scales with wave-vector as $S(k,t)\sim |k|^{-(2\alpha_\mathrm{loc}+1)}$, which depends on the local roughness exponent \cite{Lopez1997}, and not on the global one as for the FV case \cite{Barabasi1995}.

In practice, Eq.\ \eqref{eq:corr_length} allows one to compute the correlation length at a given time $t$ \cite{Barreales2020}. Indeed, we can define a correlation length $\xi_a(t)$ by means of
\begin{equation}
C_2(\xi_a(t),t)= a \, C_2(L_y/2,t)\,,
\label{eq:c2_scaling}
\end {equation}
where $a$ is a constant, typically $a \gtrsim 0.8$. In other words, the correlation length at a given time $t$ is defined as the distance along the front at which the correlation function $C_2$ takes on the $a$ fraction of its plateau-value $C_2(L_y/2,t)$. As will be shown below, the particular value of $a$ does not modify the scaling of the correlation length.

The uncertainties of the fluctuations and the correlation functions have been calculated following the jackknife procedure \cite{Young2015,Efron1982}; see also Appendix B of Ref.\ \cite{Barreales2020} for more details.

\section{Results and discussion} \label{sec:results}
In what follows, all the figures shown correspond to the dynamical evolution of the precursor layer. We refer the reader to Appendix~\ref{details}, in particular Table \ref{tab:det}, for a complete description of all the runs that we have performed.

\subsection{Front position and roughness}
\label{sec:RW}

\begin{figure}[!t]
\centering
\includegraphics[width=0.45\textwidth]{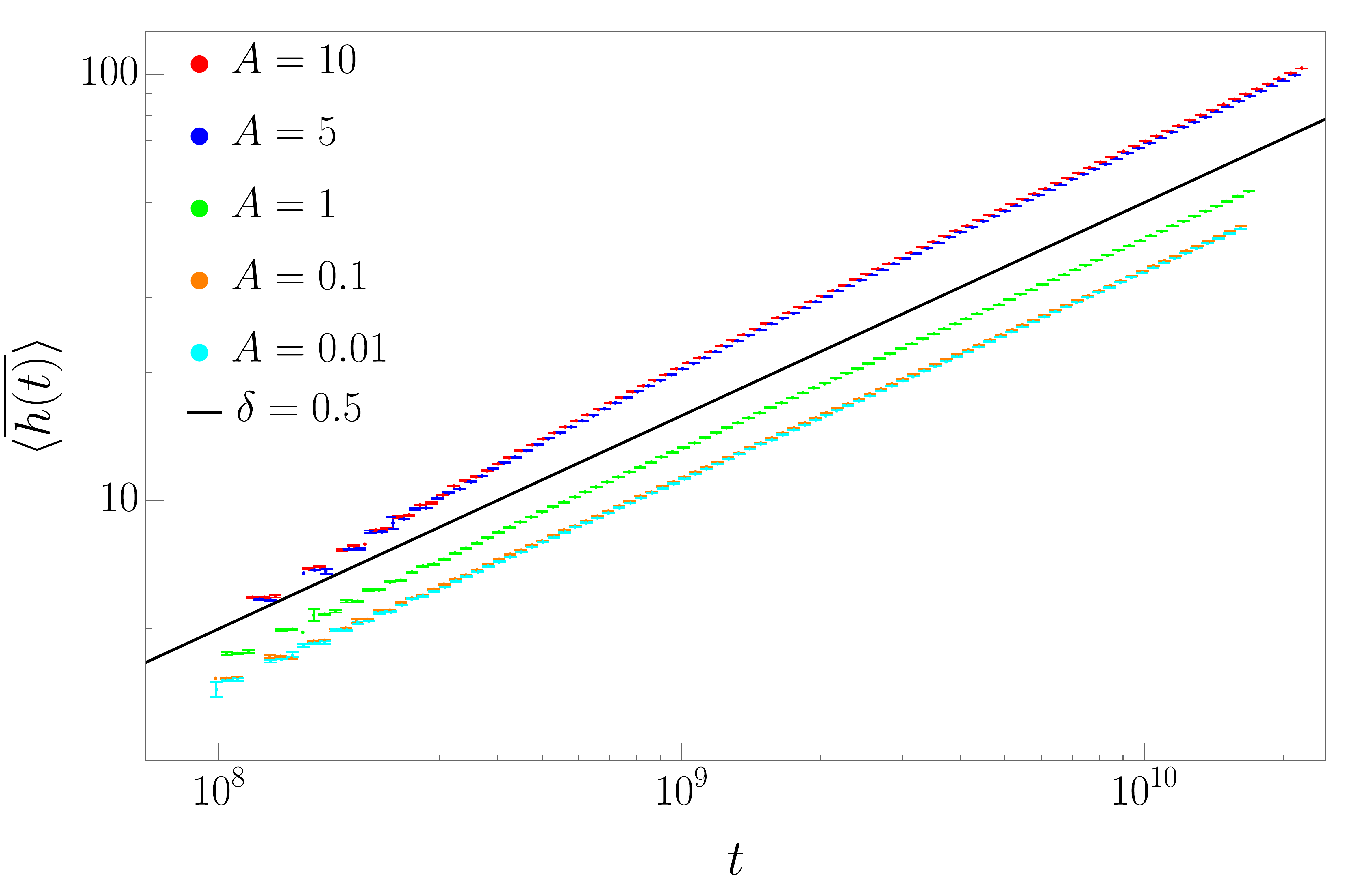}
\caption{(Color online) Average front position $\langle \bar{h}(t)\rangle$ as a function of time for $J=1$, $T=1$, $L_x =1000$, $L_y=256$, and several values of $A$. The solid black line corresponds to the reference scaling $\langle \bar{h}(t) \rangle \sim t^{1/2}$. All units are arbitrary.}
\label{fig:ht}
\end{figure}

\begin{figure}[!t]
\centering
\includegraphics[width=0.45\textwidth]{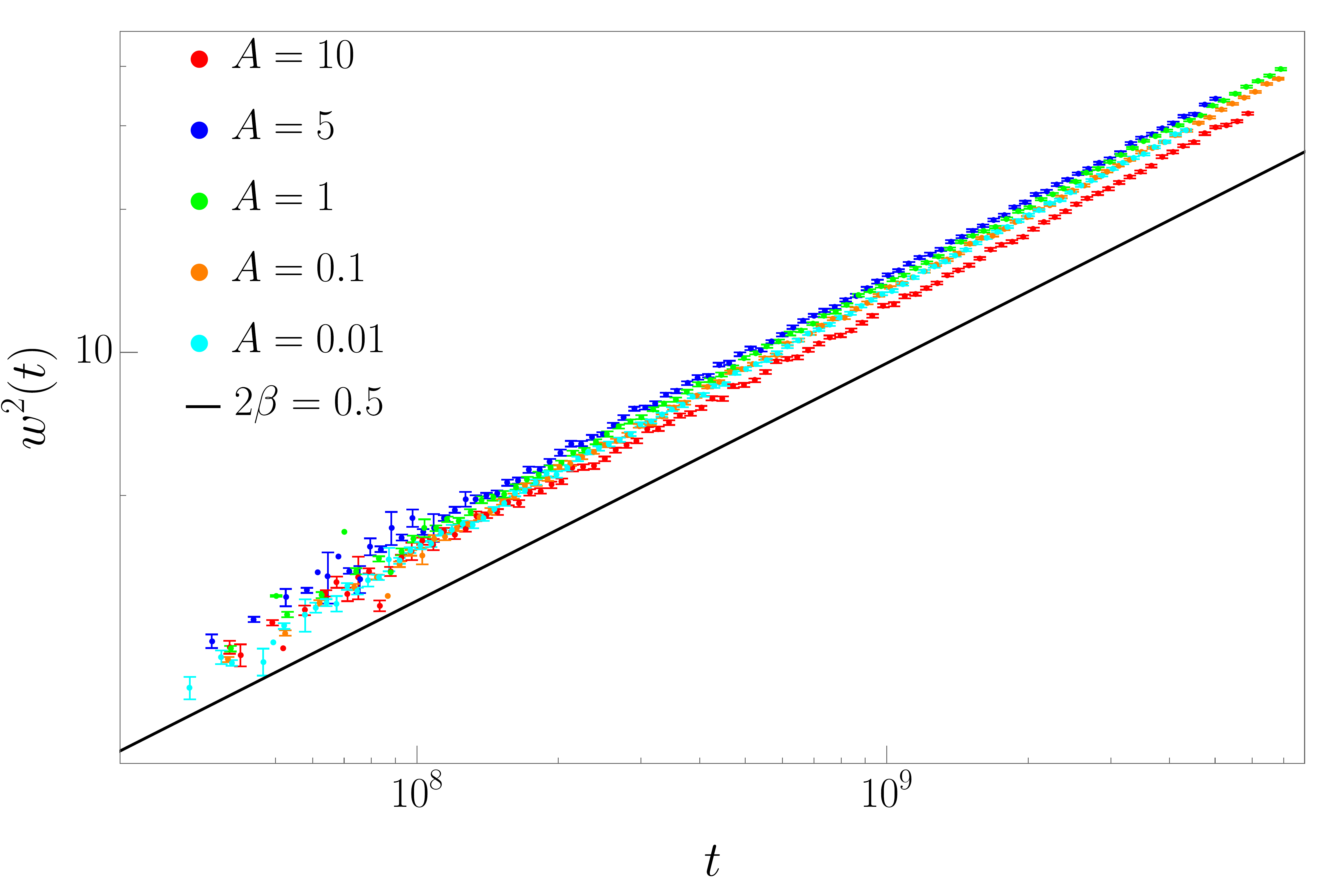}
\caption{(Color online) Squared roughness $w^2(t)$ as a function of time, as obtained for $J=1$, $T=3$, $L_x =1000$, $L_y=256$, and several values of $A$. The corresponding values of $2\beta$ are given in Table \ref{tab:beta}. As a visual reference, the solid black line corresponds to $w^2(t) \sim t^{1/2}$. All units are arbitrary.}
\label{fig:w2t}
\end{figure}

We have computed the mean front position as a function of time for two different system sizes, namely $L_y= 64$ and $L_y=256$, with $L_x=1000$ in both cases. Figure \ref{fig:ht} shows the behavior of $\langle \bar h(t)\rangle$ for five different choices of the parameters. For all values of $A$ and $T$, the mean front position grows as $\langle \bar{h}(t) \rangle \sim t^\delta$, as expected, with exponent values $\delta \approx 1/2$, as detailed by Table \ref{tab:delta} provided in Appendix~\ref{details}. As for the roughness, it scales as $w^2(t) \sim t^{2\beta}$; this behavior is shown in Fig.\ \ref{fig:w2t}.

We notice that no evidences of eventual saturation to a steady-state value \cite{Barabasi1995,Krug1997} have been observed, due to the large lattice sizes of the simulated systems. Conversely, we should also remark that we have not detected any substantial time dependence of the exponent values that will be reported below at our long times. Thus, we have avoided entering the very-long time regime explored in Ref. \cite{Abraham2002} in which the precursor film has grown so wide that diffusion is no longer able to communicate its front with the liquid reservoir efficiently, and the front evolves effectively as if there were no external driving.

\begin{figure}[!t]
\centering
\includegraphics[width=0.45\textwidth]{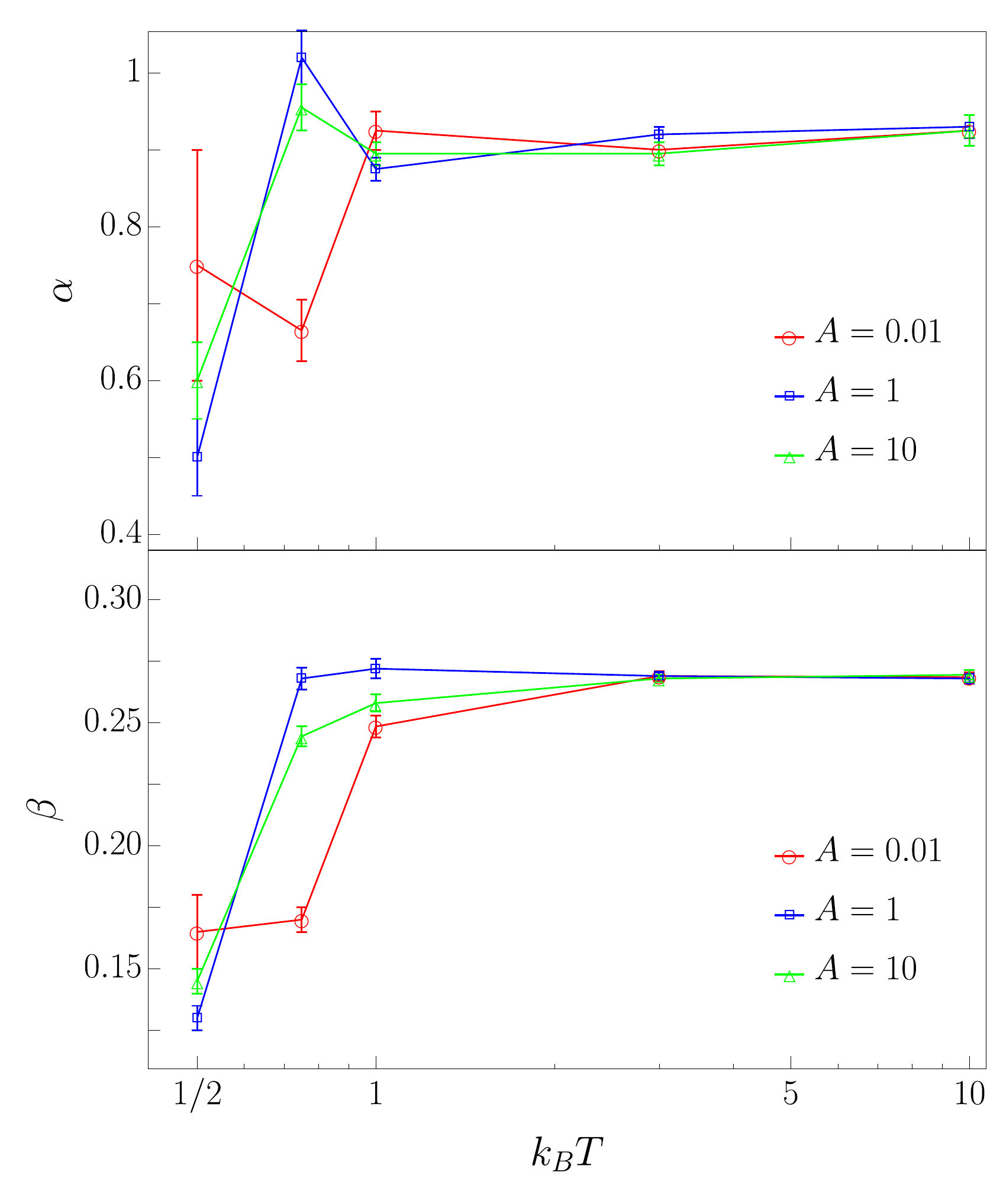}
\caption{(Color online) Values of $\alpha$ (top) and $\beta$ (bottom) taken from Table \ref{tab:a9} and Table \ref{tab:beta} vs. $T$ for $J=1$ and for $A=0.01$ (circles), $A=1$ (squares), and $A=10$ (triangles). Lines are guides to the eye. All units are arbitrary.}
\label{fig:TablaExponentes}
\end{figure}

Table \ref{tab:beta} in Appendix~\ref{details} reports the growth exponent values obtained for both layers in this study. This table indicates that the detailed value of $\beta$ depends on the physical parameters $A$ and $T$. The same data are summarized graphically in Fig.\ \ref{fig:TablaExponentes} (bottom). At high temperatures (approximately, $T > 1$), $\beta\approx 0.26$ takes essentially the same value for the precursor and supernatant layers, and does not depend on the Hamaker constant $A$. At low temperatures $(T < 1)$, the growth exponent is slightly higher for the precursor layer and seems more sensitive to the value of $A$ both for the precursor and for the supernatant layers. As a reference value in the low-temperature regime, the kMC simulations of Ref. \cite{Abraham2002} obtained $\beta\simeq1/6$ for the precursor layer using $J=1$, $A=10$, and $T=1/3$, which is compatible with our results.

Overall, Figs.\ \ref{fig:ht} through \ref{fig:TablaExponentes} already indicate a non-trivial dependence of scaling exponents with temperature, while their dependence on the Hamaker constant is comparatively much weaker. Thus, two main scaling regimes seem to exist, a low-temperature and a high-temperature one, with temperature-dependent exponents for intermediate values of $T$. As will be seen below, further exponent estimates confirm this picture.

At this point and regarding the identification of universal scaling behavior, we must note that our kMC results are unavoidably limited due to the finite size of the systems employed and the statistics assessed. From a theoretical point of view, crossover effects are known to occur very frequently in kinetic roughening processes \cite{Cuerno2004,Cuerno2007}. In the renormalization group framework \cite{Barabasi1995,Tauber2014}, these are induced by the existence of more than one attracting fixed point (FP), i.e.\ universality class, for the nonequilibrium system under study. For finite parameter values, system size, and simulation times, renormalization towards the most relevant FP may be incomplete, inducing e.g.\ effective values for the scaling exponents that are measured, which may not coincide with any of the expected universality classes. In many cases, it is at most these effective exponents which are the ones accessible experimentally \cite{Cuerno2004,Cuerno2007}. From the practical point of view of our present numerical simulations, we are unable to control the (subdominant) scaling corrections of the data due to the high correlation (in time) of the observables which we measure, and to the lack of precise theoretical predictions in the various parameter regions that we study. Without this kind of control, stronger numerical arguments regarding universal behavior can hardly be provided. In spite of these limitations, the values we report for the critical exponents show compatibility in a wide range of parameters, from a statistical point of view (differences at most less than two standard deviations). Moreover, the long simulated times gives us some confidence that the contribution of the subdominant terms can be safely neglected.

\subsection{Height-difference correlation function: computation of $\alpha$ and $z$ exponents}
\label{sec:hdcf}
In this section we will study the height-difference correlation function.

\begin{figure}[!t]
\centering
\includegraphics[width=0.45\textwidth]{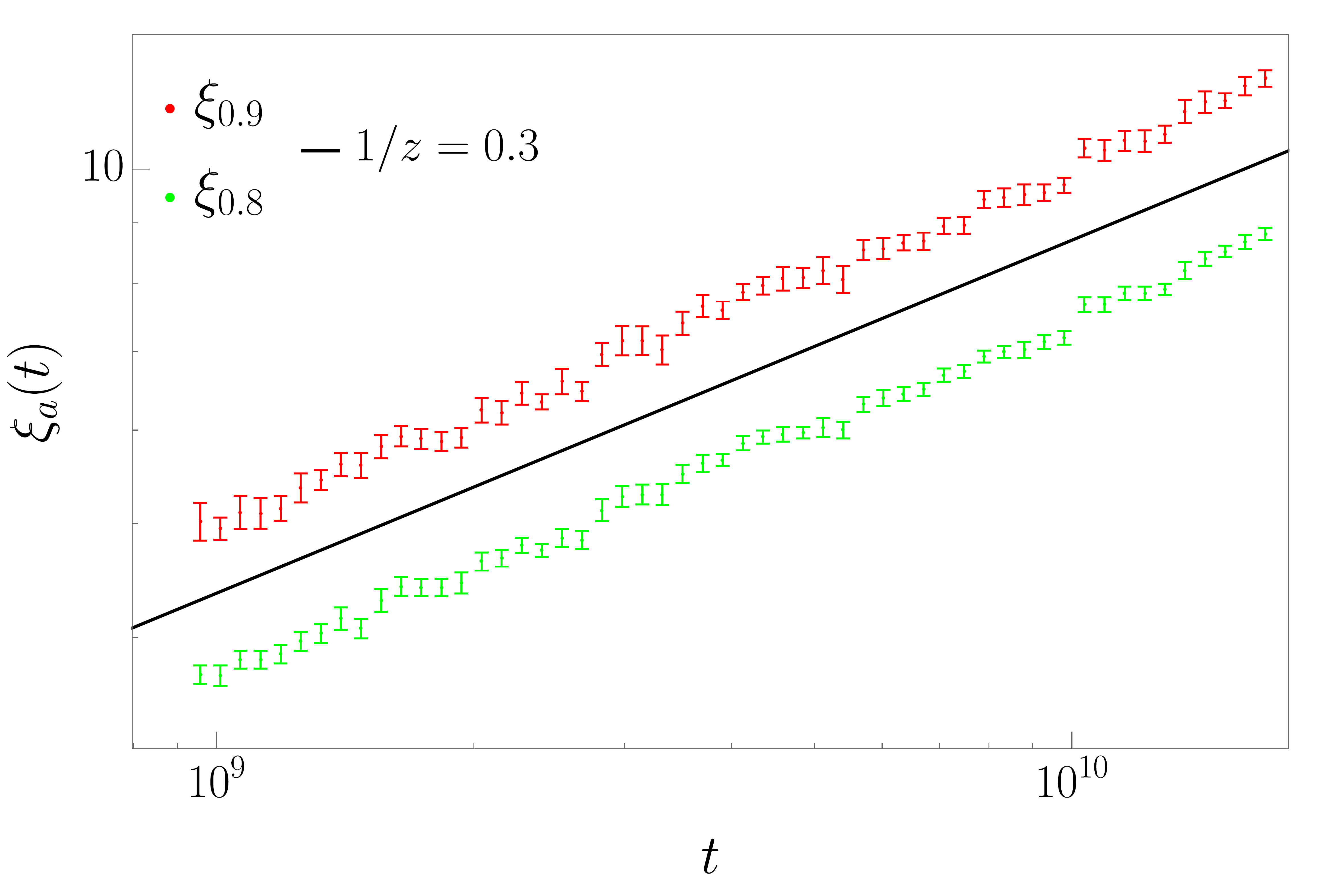}
\caption{(Color online) Estimates $\xi_{0.8}(t)$ and $\xi_{0.9}(t)$ as functions of time, obtained for $J=1$, $T=1$, $A=1$ and $L_x=1000$. As a visual reference, the solid black line corresponds to $\xi(t) \sim t^{0.3}$. All units are arbitrary.}
\label{fig:dc}
\end{figure}

As described in Sec.\ \ref{observables}, the correlation length at a given time $t$, $\xi(t)$, can be estimated from the plateau of the $C_2(r,t)$ curves at large enough $r$ for different values of $a$. From Eq.\ \eqref{eq:correlation_length}, the double logarithmic plots of these correlation lengths as functions of time should fit straight lines whose slopes are the same $1/z$ exponent. Figure \ref{fig:dc} shows $\xi_a(t)$ vs $t$ in log-log plots for the precursor layer, calculated for $a = 0.8$ and $a = 0.9$, with an exponent $1/z\sim 0.3$.

On the other hand, Eq.\ \eqref{eq:corr_length} yields $C_2(r,t)=\xi^{2\alpha}(t)$ for $r \gg \xi(t)$. Thus, the $\alpha$ exponent may be calculated from the slope of the best-fit lines in a $C_2(r,t)$ versus $\xi(t)$ log-log plot; for simplicity, we have plotted the correlation functions evaluated at $r = L_y/2$.
\begin{figure}[!t]
\centering
\includegraphics[width=0.45\textwidth]{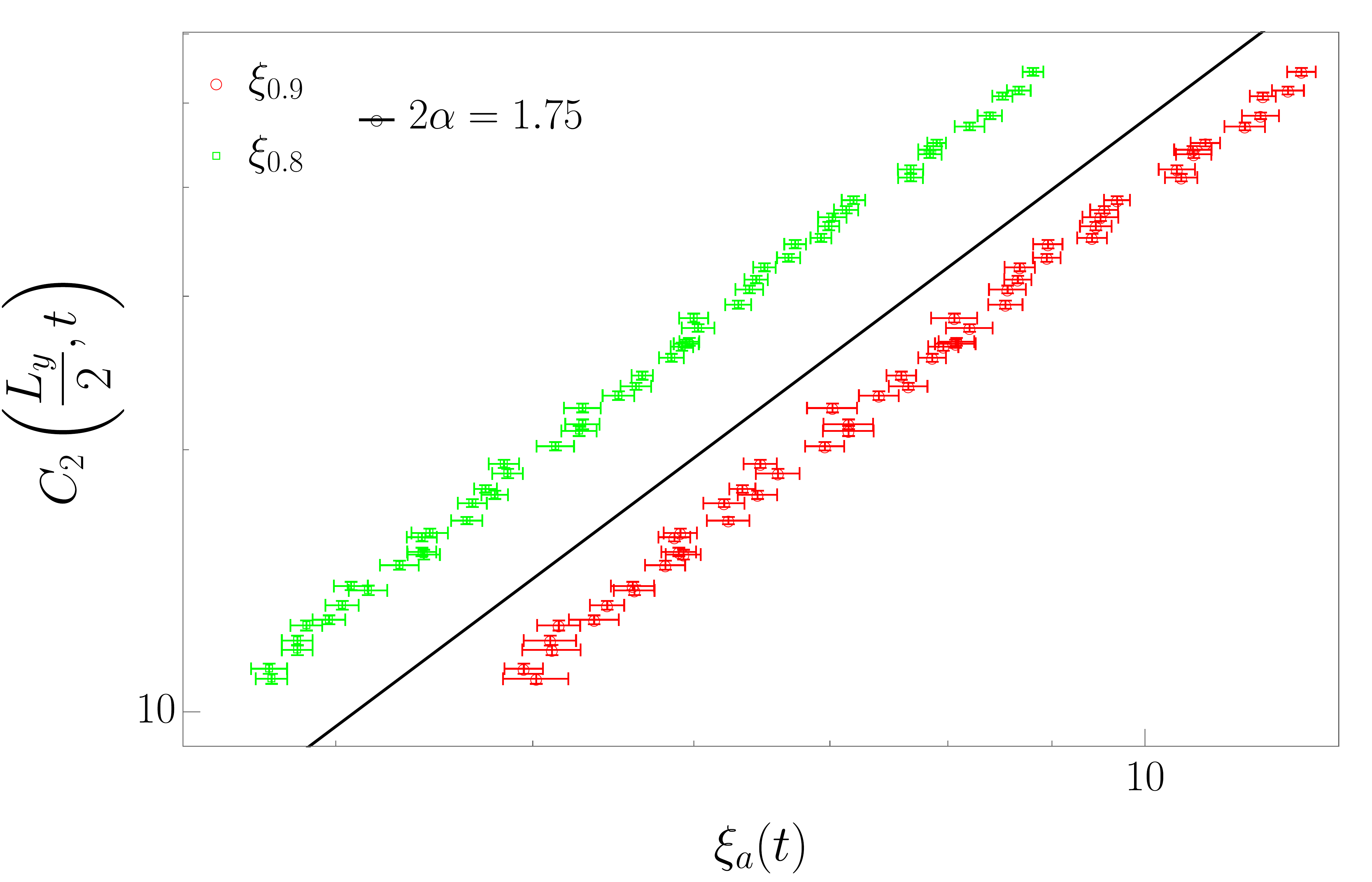}
\caption{(Color online) Height-difference correlation function $C_2\left(L_y/2,t\right)$ versus $\xi_{0.8}(t)$ and $\xi_{0.9}(t)$ at different times. Conditions are $J=1$, $T=1$, $A=1$, $L_x=1000$, and $L_y=256$. As a visual reference, the solid black line corresponds to $C_2\left(L_y/2,t\right) \sim t^{1.75}$. All units are arbitrary.}
\label{fig:plvsdc}
\end{figure}
In Fig.\ \ref{fig:plvsdc} we plot $C_2(L_y/2,t)$ against $\xi_a(t)$ for the precursor layer and the same values of $a$ with $2 \alpha \sim 1.75$.

The complete set of $1/z$ and $2\alpha$ exponents, calculated for $a = 0.8$ and $a=0.9$, are given in Tables \ref{tab:a8} and \ref{tab:a9}, respectively, in Appendix~\ref{details}. From these data, one may easily check that the expected scaling relation $\alpha=\beta z$ holds. A representative choice of these results for the $\alpha$ and $\beta$ exponents is displayed graphically in Fig.\ \ref{fig:TablaExponentes}.

Analogously to what was anticipated above regarding $\beta$, while the dependence of $\alpha$ and $z$ with the Hamaker constant is relatively marginal, their dependence with temperature is much more substantial and similarly suggests a transition from a low-temperature to a high-temperature regime, with $T$-dependent exponents for intermediate temperatures around $T=1$. In general, notice that both $\alpha$ and $z$ change quite abruptly with $T$ from their low-$T$ values into $\alpha\approx0.9$ and $z\approx3.4$ for the high-$T$ regime.

\begin{figure}[!t]
\centering
\includegraphics[width=0.47\textwidth]{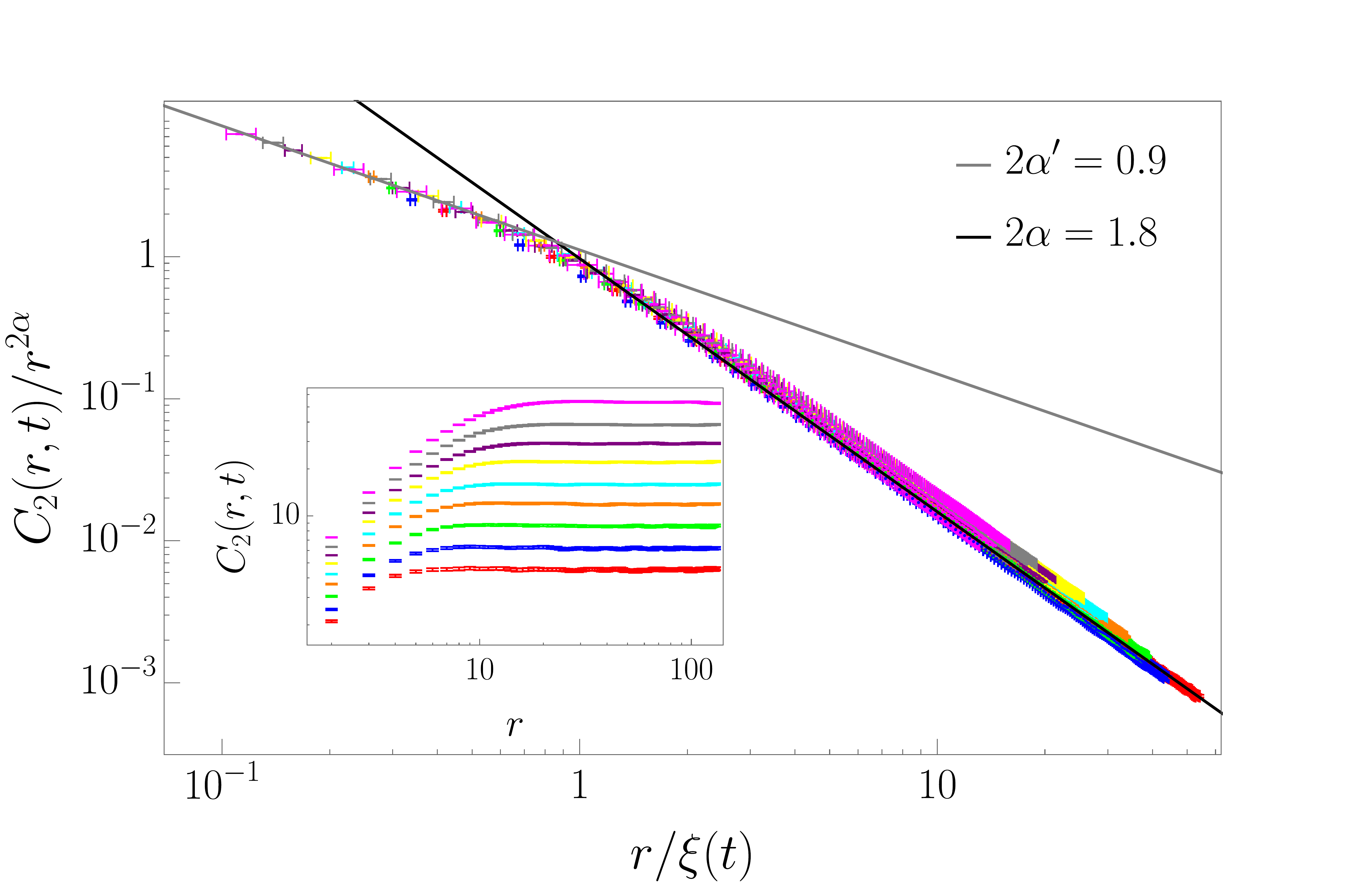}
\caption{(Color online) Data collapse of the height-difference correlation function obtained for different values of time, for $J=1$, $T=1$, $A=1$, $L_x=1000$, and $L_y=256$, using $\alpha=0.9$. The curve onto which collapse occurs is the function $g(r/\xi(t))$ of Eq.\ \eqref{eq:corr_length}, the solid black line representing the theoretical behavior for large $u$, $g(u) \sim u^{-2\alpha}$,  and the solid gray line representing the behavior for small $u$, $g(u) \sim u^{-2\alpha'}$ (See Tables \ref{tab:a8} and \ref{tab:ap} in the Appendix). All units are arbitrary. Inset: Height-difference correlation function as a function of $r$ for times increasing from 20 to 100 bottom to top at regular intervals.}
\label{fig:c2}
\end{figure}
\subsection{Anomalous scaling of the height-correlation function}
Values of the global roughness exponent $\alpha \lesssim 1$, as obtained here for $T \gtrsim 1/2$, speak of large fluctuations in the front position. In our present case, they turn out to be associated with intrinsic anomalous scaling. An indication of this fact is apparent in the inset of Fig.\ \ref{fig:c2}. Indeed, the fact that $C_2(r,t)$ curves obtained for different times displace systematically with time and do not overlap is a landmark behavior of anomalous scaling which, in principle, can be originated by different causes, large values of $\alpha$ (so-called superroughening) being one of them \cite{Lopez1997}. In our case, it stems from the fact that $\alpha_{\rm loc}\neq\alpha$, there being two independent roughness exponents. This is unambiguously shown in the main panel of Fig.\ \ref{fig:c2}, which displays a consistent data collapse of the height-difference correlation function according to Eq.\ \eqref{eq:corr_length}, for a representative parameter choice.

If the scaling behavior were of the standard FV type, the scaling function $g(u)$ would be $u$-independent at small arguments $u \ll 1$; on the contrary, our data agree with a scaling law of the form $g(u) \sim u^{-2\alpha'}$, with $2\alpha' \approx 0.9$, so that $\alpha_\mathrm{loc}\approx 0.45$ while $\alpha=0.89$, implying the occurrence of intrinsic anomalous scaling \cite{Lopez1997}. Analogous behavior is found for other parameter choices, see Table \ref{tab:ap} in Appendix~\ref{details} for specific exponent values.

\begin{figure}[!t]%
    \centering
    \includegraphics[width=0.45\textwidth]{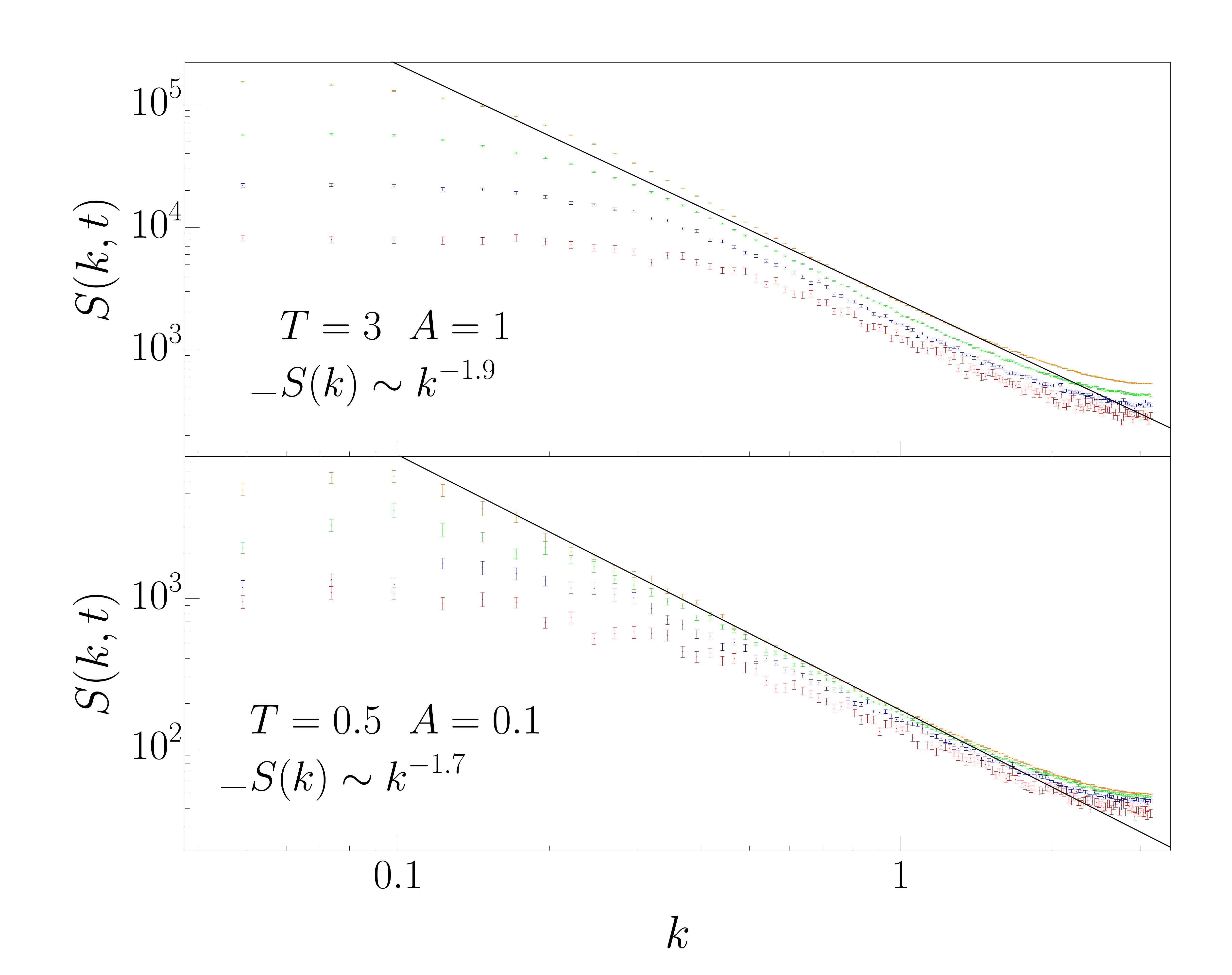}
    \label{subfig:structureFactor}
    \caption{(Color online) Structure factor calculated for the precursor layer at (a) $J=1$, $T = 0.5$, $A=0.1$ and (b) $J=1$, $T = 3$, $A=1$, for times increasing bottom to top in both panels. The scaling behavior at fixed time is $S(k,t) \sim |k|^{-(2\alpha_{\rm loc}+1)}$, where $\alpha_{\rm loc}$ has been evaluated as $\alpha-\alpha'$, see Tables \ref{tab:a8} and \ref{tab:ap} in the Appendix. The power laws represented by the solid lines are indicated in the corresponding legends. All units are arbitrary.}%
    \label{fig:structureFactor}%
\end{figure}

As indicated above, the anomalous shift of the height-difference correlation function curves with increasing time illustrated in the inset of Fig.\ \ref{fig:c2} could be alternatively induced by a mere large roughness exponent. However, the collapse of the same data in Fig.\ \ref{fig:c2} with $\alpha'\neq 0$ unambiguously identifies the origin of this behavior, rather, as intrinsic anomalous scaling. It is still worth examining further the system behavior by means of the structure factor. This is because it allows us to reinterpret previous results from Ref. \cite{Abraham2002} based on this observable and because the behavior we presently obtain for $S(k,t)$ further confirms the occurrence of intrinsic anomalous scaling. In Fig.\ \ref{fig:structureFactor} we show the structure factor, calculated for various times and two representative temperatures, namely $T = 0.5$ and $T = 3$. Note that the $S(k,t)$ curves also shift upwards systematically with time in agreement with Eq.\ \eqref{eq:Sk}, this being another landmark of intrinsic anomalous scaling \cite{Lopez1997}. As a result, as noted above $S(k,t)\sim |k|^{-(2\alpha_\mathrm{loc}+1)}$ for long enough times~\cite{Lopez1997}, so that the roughness exponent that can be read off from the power-law behavior of $S(k,t)$ seen in Fig.\ \ref{fig:structureFactor} is $\alpha_{\rm loc}$, and not $\alpha$. This applies in particular to the results in Ref. \cite{Abraham2002}: while the systematic time shift of the structure factor can be unambiguously seen in Fig.\ 3(a) of that paper, this issue on the interpretation of the scaling exponents was overlooked there. Hence, we interpret that the low-temperature roughness exponent obtained in Ref.\ \cite{Abraham2002} was the local, rather than the global one.

\subsection{Additional universal properties of the fronts: probability distribution function and front covariance}

\begin{figure}[!htbp]
\centering
\includegraphics[width=0.45\textwidth]{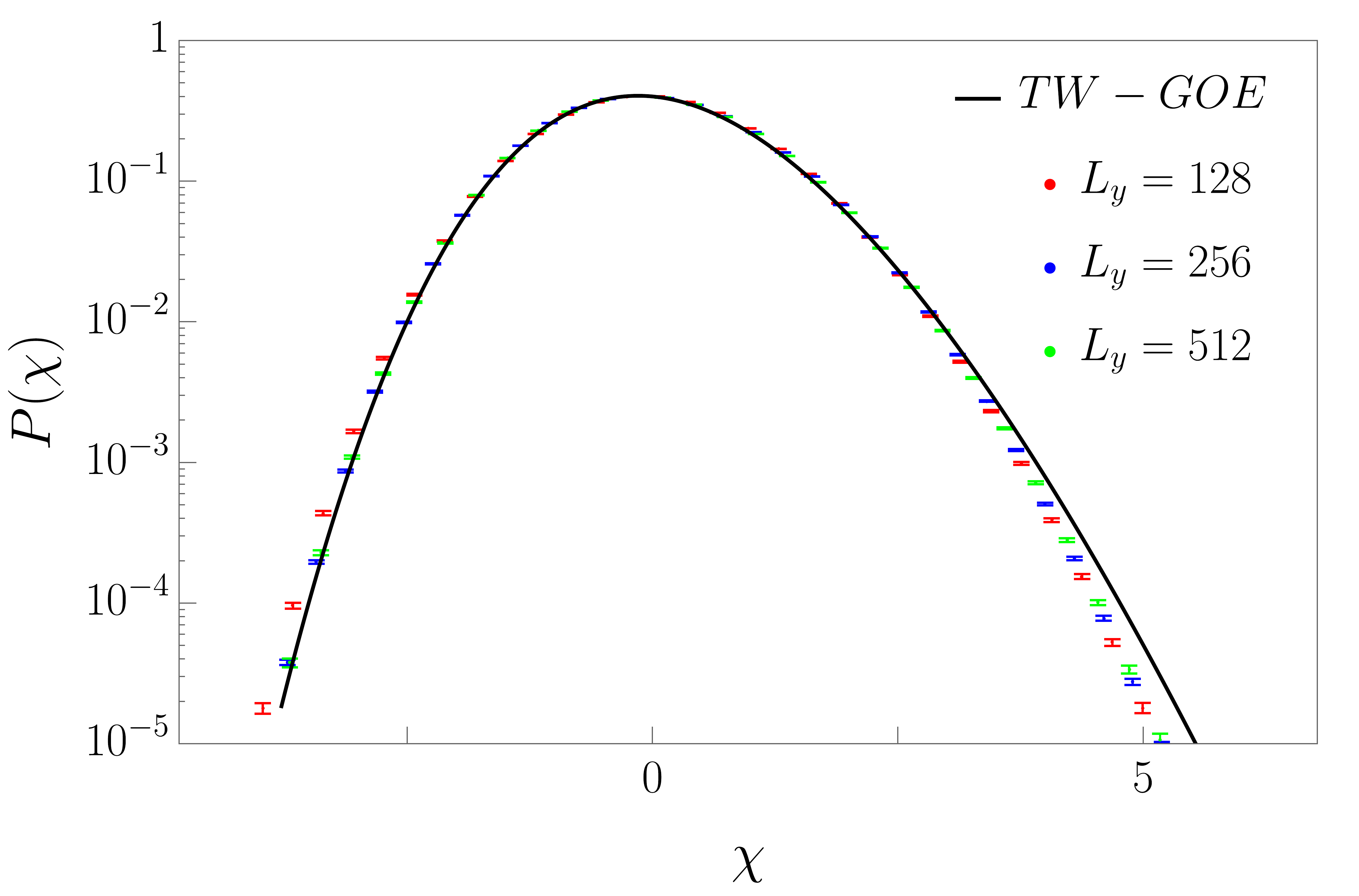}
\caption{(Color online) Fluctuation histograms calculated according to Eq.\ \eqref{eq:fluctuations} for $J=1$, $A = 1$, $T=1$, and several system sizes, as indicated in the legend. The solid line corresponds to the GOE Tracy-Widom distribution. All units are arbitrary.}
\label{fig:histograma}
 \end{figure}

As noted in the Introduction, recent developments on surface kinetic roughening, mostly in the context of KPZ scaling, have shown that universal behavior goes beyond the values of the critical exponents for many important universality classes. Specifically, by normalizing the fluctuations of the front around its mean by their time-dependent amplitude as
\begin{equation}
	\label{eq:fluctuations}
	\chi(y,t) = \frac {h(y,t) - \bar{h}(t)}{t^{\beta}} ,
\end{equation}
the probability density function (PDF) of these $\chi$ random variables becomes time-independent and is shared by all members of the universality class \cite{Kriecherbauer2010,HalpinHealy2015,Carrasco2016,Takeuchi2018,Carrasco2019}.
This effect is demonstrated in Fig.\ \ref{fig:histograma}, which shows that the PDF corresponding to various system sizes collapse into a single curve in a range of front fluctuations.\footnote{In particular,
using the formulas given in Eqs.\ \eqref{Eq:Skewness} and \eqref{Eq:Kurtosis},
we find the skewness and excess kurtosis for the PDF in Fig.\ \ref{fig:histograma} to be $S =0.221(3)$, $K = 0.239(5)$ for $L_y = 128$, $S=0.236(2)$, $K =0.249(1)$ for $L_y = 256$, and $S = 0.264(2)$, $K = 0.239(4)$ for $L_y = 512$, see e.g.\ \cite{Kriecherbauer2010,HalpinHealy2015} for definitions of $S$ and $K$. These values suggests that, while $K$  seems to change little with system size, $S$ does increase with $L_y$. For reference, in Fig.~\ref{fig:histograma} we have plotted the Tracy-Widom distribution for the largest eigenvalue of a random matrix in the Gaussian orthogonal ensemble (TW-GOE) \cite{Kriecherbauer2010,HalpinHealy2015}, whose precise skewness and excess kurtosis values are $S=0.29346452408$ and $K=0.1652429384$, respectively \cite{Bornemann2010}.}

It has been shown that the TW-GOE PDF describes accurately the fluctuations of fronts in the KPZ universality class when periodic boundary conditions are employed, as in our simulations; see additional references e.g.\ in Ref.\ \cite{Barreales2020}. In view of the fact that the kinetic roughening of our kMC fronts is intrinsically anomalous (while it is standard FV type for the KPZ equation \cite{Kardar1986}) and with non-KPZ exponents, the agreement of our numerical PDF with the TW distribution for $|\chi| \lesssim 2.5$ is unexpected.

The front covariance $C_1(r,t)$, defined in Eq.~\eqref{eq:correlation_1}, also exhibits KPZ behavior. In general, this function is expected to behave as
\begin{equation}
    C_1(r,t)=a_1 \, t^{2\beta} f\left(a_2 r/t^{1/z} \right) ,
   \label{eq:airy1}
\end{equation}
where $f(u)$ is an universal function and $a_1$ and $a_2$ are non-universal constants \cite{Alves2011,Oliveira2012,Nicoli2013} to be computed in our simulations. For the specific case of a system described by the one-dimensional KPZ equation with periodic boundary conditions, $f(u)\equiv \mathrm{Airy_1}(u)$, where $\mathrm{Airy}_1(u)$ denotes the covariance of the Airy$_1$ process \cite{Bornemann2009,HalpinHealy2015,Takeuchi2018}. The computation of $a_1$ and $a_2$ follows the same procedure as in Ref.\ \cite{Barreales2020}.

\begin{figure}[!t]
\centering
\includegraphics[width=0.45\textwidth]{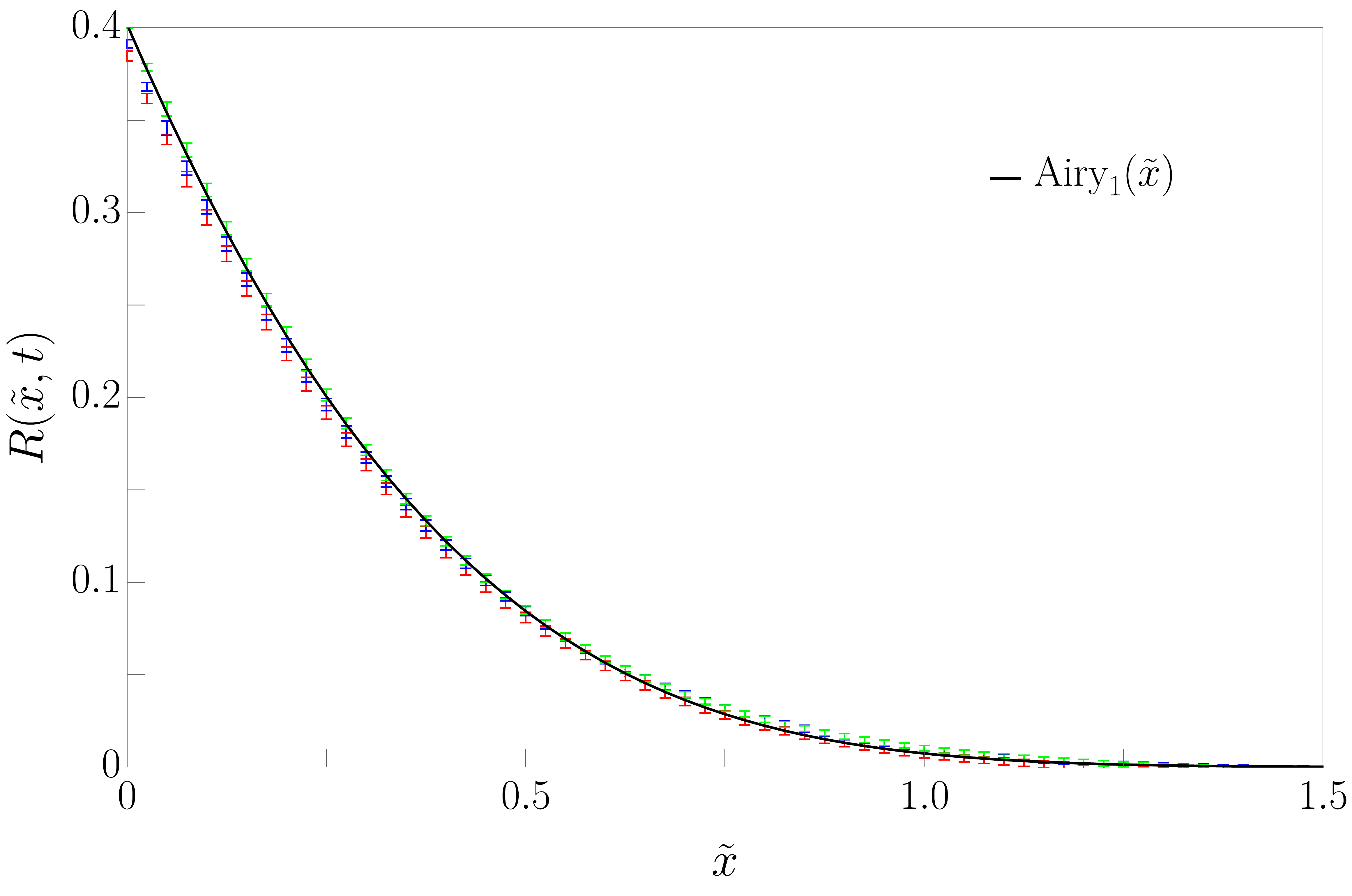}
\caption{(Color online) $R(\tilde{x},t) \equiv \frac{C_1\left(\tilde{x} t^{1/z}/a_2\right)}{a_1t^{2\beta}}$ versus $\tilde{x}\equiv a_2 r/t^{1/z}$ for $t=60$, $80$, and $100$, calculated for parameters as in Fig.\ \ref{fig:histograma}, using $1/z=0.32 $, $2\beta=0.544 $, $a_1= 1.834\times10^{-4}$, and $a_2=8.985\times 10^{-3}$. The solid line corresponds to the exact $\mathrm{Airy}_1(\tilde{x})$ function. All units are arbitrary.}
\label{fig:c1}
\end{figure}

Figure \ref{fig:c1} shows the collapsed height covariance functions $C_1(\tilde xt^{1/z}/a_2)/(a_1t^{2\beta})$ vs $\tilde x$ for several time intervals. This figure shows that the universal behavior implied by Eq.\ \eqref{eq:airy1} holds with $f(u)=\mathrm{Airy_1}(u)$, even if the exponent values are not those of 1D KPZ  and in spite of the fact that the scaling is intrinsically anomalous. We must note that this agreement deteriorates as the temperature decreases, so that for $T \lesssim T_* = 3/4$ our numerical two-point statistics substantially deviates from the Airy$_1$ form.

\section{Continuum equation in the high-temperature regime}
\label{sec:eq}

The 1D KPZ behavior of the front fluctuations that we have just discussed can be partly rationalized by considering a continuum equation which is expected to describe the large-scale properties of the spreading fronts that we are addressing here. Specifically, it was observed in Ref. \cite{Abraham2002} that matter transport in the lattice gas model is dominated by diffusion of particles on the supernatant layer from the reservoir to the spreading front, and by diffusion of holes in the precursor layer from its edge back to the reservoir. Based on this observation, a continuum model was put forward in which such transport processes were coupled with the motion of the precursor edge. Finally, under various approximations a single stochastic continuum equation was put forward for $h(y,t)$ in \cite{Abraham2002}, which was then favorably compared with the kMC simulations reported in that reference. For our present discussion, we simplify such an equation to a form expected to be relevant to the high-$T$ regime of our present simulations. Specifically, we consider
\begin{eqnarray}
\partial_t \hat{h}_k(t) & = & -\nu |k|^3 \hat{h}_k(t) + \frac{\lambda}{\sqrt{t}} \hat{\mathcal{F}}_k[(\partial_y h)^2] + \hat{\eta}_k(t) , \nonumber \\
 & & \langle \hat{\eta}_k(t) \hat{\eta}_{k'}(t') \rangle = D \delta_{k+k'} \delta(t-t'),
\label{eq:eom}
\end{eqnarray}
where $\hat{h}_k(t)$ is the space Fourier transform ($\hat{\mathcal{F}}_k$) of the $h(y,t)$ front of the precursor layer, $\nu>0, \lambda$ are parameters related with those of the original moving boundary problem \cite{Abraham2002} (diffusion coefficients of particles and holes, temperature, etc.), and $\hat{\eta}_k(t)$ is the Fourier transform of zero-average, Gaussian white noise $\eta(y,t)$. Equation \eqref{eq:eom} is a particular case of the more complex interface equation derived in \cite{Abraham2002}, and retains key features which are relevant to the high-$T$ regime of our kMC simulations. On the one hand, the nonlinear term of Eq.\ \eqref{eq:eom} is precisely the KPZ nonlinearity, which directly brings KPZ scaling into the discussion. The peculiarity here is that, rather than being time-independent as in the standard KPZ equation, the coupling of the nonlinear term is inversely proportional to $t^{1/2}$. Physically, this fact originates in the diffusive coupling between the fluid reservoir and the front of the precursor film. Akin to the Lucas-Washburn law in fluid imbibition systems \cite{Alava2004}, it is such a diffusive coupling which leads to the expected motion of the average front $\langle \bar{h}(t) \rangle \sim t^{\delta}$ with $\delta\approx1/2$. In turn, such a growth law implies an average front velocity $V(t) = d\langle \bar{h}(t) \rangle/dt \approx 1/t^{1/2}$, and indeed in the KPZ equation $V(t)$ is proportional to the coupling of the nonlinear term \cite{Kardar1986,Barabasi1995,Krug1997}. Further distinctive features of Eq.\ \eqref{eq:eom} are the non-local linear term with parameter $\nu$ and the non-conserved noise \cite{Abraham2002}: the former stems from effective surface tension of the front (arising from evaporation/condensation of particles and holes there), combined with non-local (geometric shadowing) effects which are well-known in diffusion-limited growth systems \cite{Nicoli2008}. In the present system, this nonlocal linear term implies that a depression in the shape of the precursor front will be filled by liquid preferentially over the growth of a protrusion, and this tends to smooth out fluctuations of the front shape. Finally, as long as the precursor edge does not decouple effectively from the reservoir at very long times in the sense noted in Sec.\ \ref{sec:RW}, the front is subjected both to conserved and non-conserved noise \cite{Abraham2002}, the latter being more relevant than the former at large scales.

As noted above, Eq.\ \eqref{eq:eom} is a simplification of the highly complex front equation derived in Ref.\ \cite{Abraham2002}, which is intended to address the high-$T$ regime of the spreading process unveiled by our kMC simulations. In order to study this nonlinear equation, we resort to numerical simulations of it in which we employ a pseudospectral scheme with integrating factor, previously described in \cite{Nicoli2008}. We set $\nu=\lambda=D=1$ and $L_{y}=2^{13}=8192$, with lattice and time spacings $\Delta x=1$ and $\Delta t=10^{-2}$. We have carried out $N=504$ iterations of the code to obtain the averages we are going to discuss in what follows.

\begin{figure}[!t]
\centering
\includegraphics[width=0.45\textwidth]{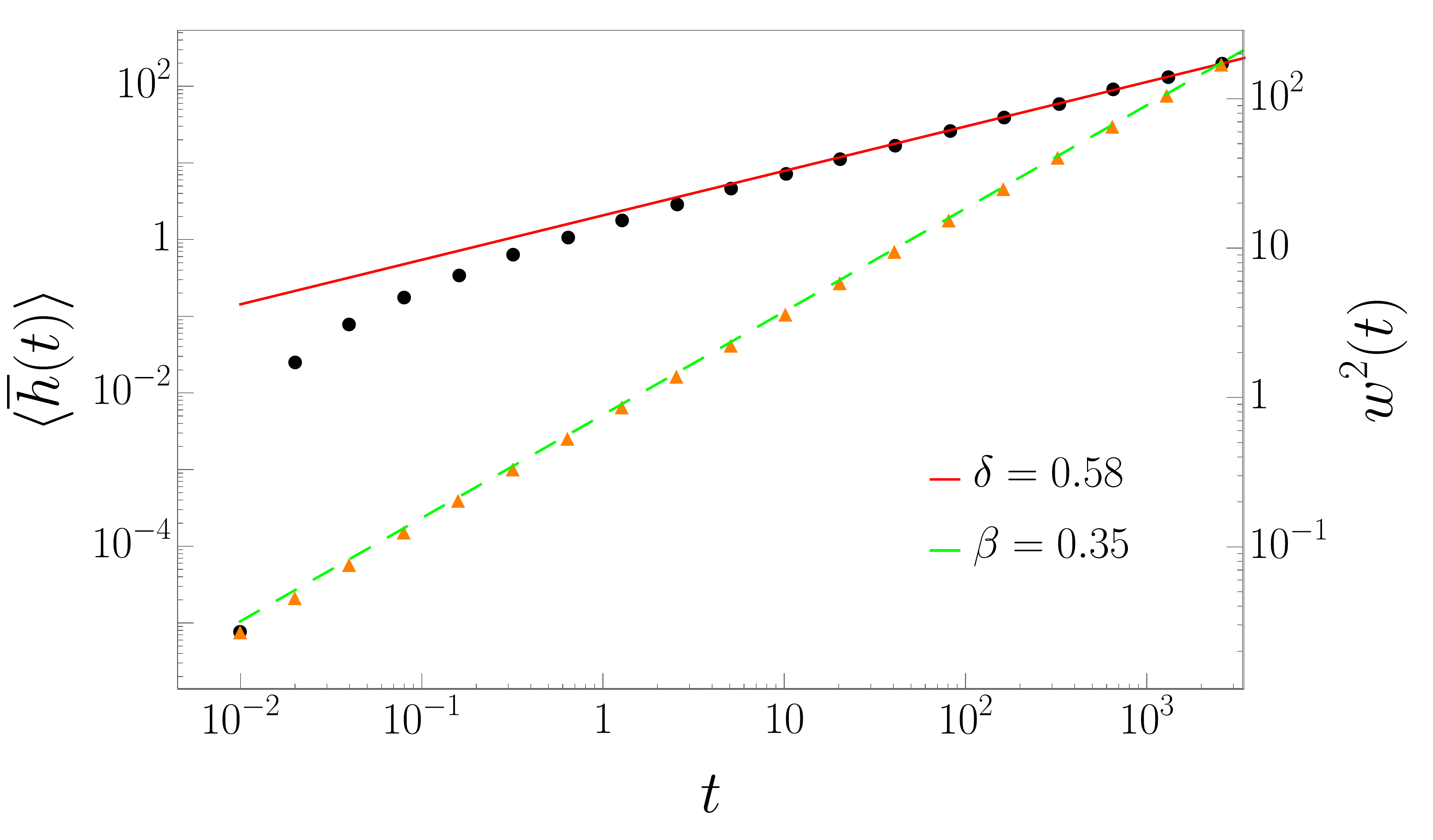}
\caption{(Color online) Average front position (left vertical axis, circles) and squared roughness (right vertical axis, triangles) as functions of $t$, as obtained from numerical simulations of Eq.\ \eqref{eq:eom} for $\nu=\lambda=1$. The solid red line correspond to power law $\langle \bar{h}(t) \rangle \sim t^{\delta}$ and the dashed green line to $w^2(t) \sim t^{2\beta}$ with exponent values as in the legend. All units are arbitrary.}
\label{fig:eq_1}
\end{figure}
\begin{figure*}[!t]
\begin{center}
\includegraphics[width=\textwidth]{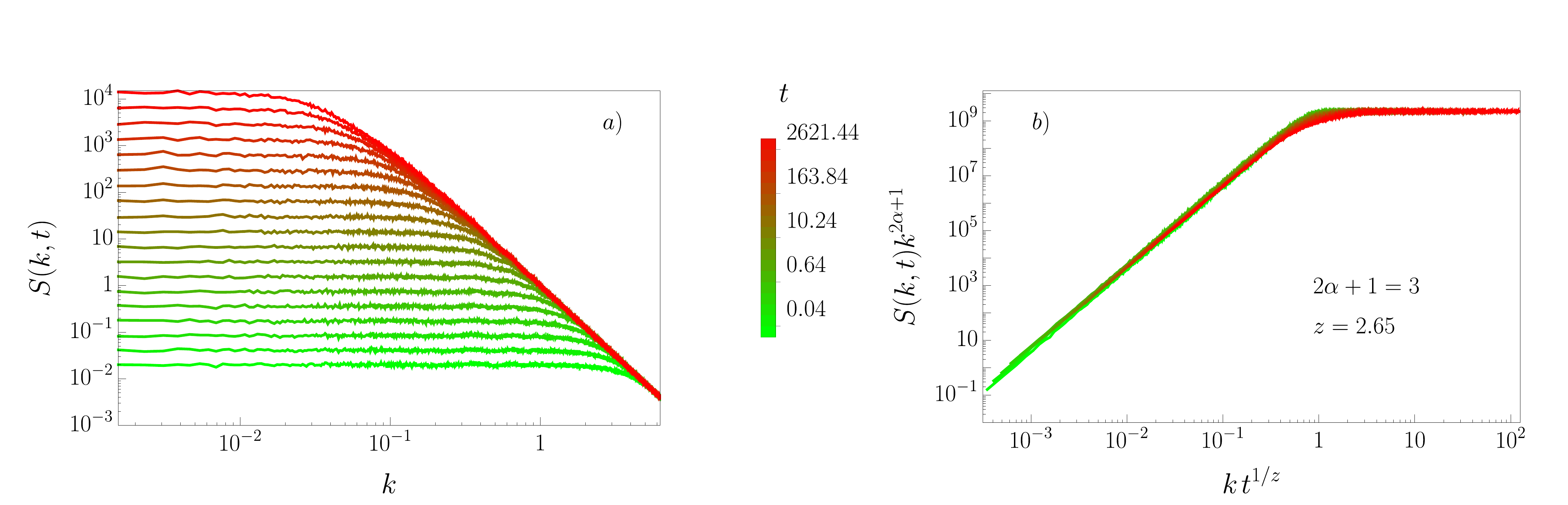}
\caption{(Color online) (a) Structure factor of the front, $S(k,t)$, as a function of wave-vector modulus $k$ for increasing times as indicated by the color scale, from numerical simulations of Eq.\ \eqref{eq:eom} using parameters as in Fig.\ \ref{fig:eq_1}. (b) Collapse of the data of panel (a) following Eq.\ \eqref{eq:Sk}. The $u$-independent behavior at large $u=kt^{1/z}$ indicates standard FV scaling with $\alpha=\alpha_{\rm loc}\simeq 1$. All units are arbitrary.}
\label{fig:eq_2}
\end{center}
\end{figure*}
Results for the average front position and front roughness are provided in Fig.\ \ref{fig:eq_1}. The time evolution of the average front position turns out to be of the expected power-law form, $\langle \bar{h}(t) \rangle \sim t^{\delta}$, with $\delta\simeq 0.58$. Hence, the average velocity of the front decays with time, as expected due to the time-decreasing KPZ coupling in Eq.\ \eqref{eq:eom}. However, the amplitude of front fluctuations does increase with time as indicated by the power-law behavior of the roughness, $w(t) \sim t^{\beta}$ with $\beta \simeq 0.35$, seen in Fig.\ \ref{fig:eq_1}. For reference, recall that for the 1D KPZ universality class one has $\beta_{\rm KPZ}=1/3$ \cite{Kardar1986,Barabasi1995,Krug1997}. However, we believe closeness of the growth exponent to the KPZ value is coincidental, as it does not occur for other exponents. This is confirmed by Fig.\ \ref{fig:eq_2}, which shows the structure factor $S(k,t)$ for different times. Note that the large-$k$ behavior of the various curves is time-independent and exhibits no anomalous shift with time, in contrast with the kMC results shown e.g.\ in Fig.\ \ref{fig:structureFactor}. Hence, scaling behavior is standard Family-Vicsek type for Eq.\ \eqref{eq:eom}, as further confirmed by the data collapse performed in Fig.\ \ref{fig:eq_2}(b) according to Eq.\ \eqref{eq:Sk}, whereby the $u$-independent behavior obtained for large $u$ in the numerical scaling function implies $\alpha=\alpha_{\rm loc}\simeq 1$. Hence, neither $\alpha=1$ nor the $z\simeq 2.65$ value implied by the collapse, are anywhere close to the 1D KPZ values ($\alpha_{\rm KPZ}=1/2$ and $z_{\rm KPZ}=3/2$), in spite of the fact that their ratio $\beta=\alpha/z \simeq 0.37$ is not so far from $\beta_{\rm KPZ}$.

Recalling the results of our kMC simulations at high $T$ $(\alpha_{\rm kMC}\simeq 0.90,\beta_{\rm kMC}\simeq 0.26,z_{\rm kMC}\simeq 3.3,\delta_{\rm kMC}\simeq 0.47)$, the exponent values predicted by Eq.\ \eqref{eq:eom} $(\alpha_{\rm eqn}\simeq 1,\beta_{\rm eqn}\simeq 0.37,z_{\rm eqn}\simeq 2.65,\delta_{\rm eqn}\simeq 0.58)$ are somewhat off, especially $z$ and hence $\beta$, while $\delta$ and specially $\alpha$ compare better.

\begin{figure*}[!t]
\begin{center}
\includegraphics[width=\textwidth]{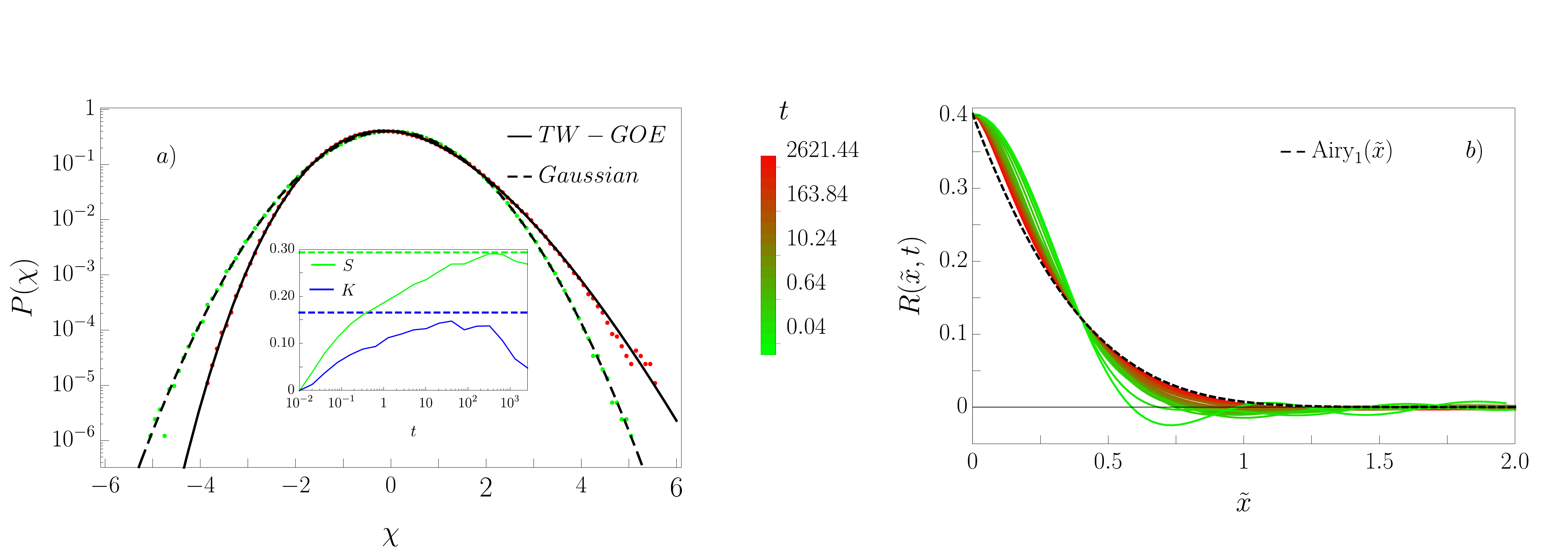}
\caption{(Color online) (a) PDF of front fluctuations as obtained from numerical simulations of Eq.\ \eqref{eq:eom} using parameters as in Fig.\ \ref{fig:eq_1}, for $t = 10^{-2}$ (green points) and for $t = 2621.44 = 2^{18}\cdot 10^{-2}$ (red points). The dashed line shows a Gaussian distribution while the solid line corresponds to the exact TW-GOE form. Inset: time evolution of the skewness and excess kurtosis corresponding to the same set of simulations as the main panel. The TW-GOE values are the abscissas of the corresponding horizontal lines and are shown for reference. (b) Data collapse according to Eq.\ \eqref{eq:airy1} for the same simulations as in panel (a). For reference, the exact Airy$_1$ covariance is shown as a dashed line. All units are arbitrary.}
\label{fig:eq_3}
\end{center}
\end{figure*}

We assess the front statistics predicted by the continuum model, whose behavior is also similar to that obtained for the high-$T$ kMC case. Results from numerical simulations of Eq.\ \eqref{eq:eom} are provided in Fig.\ \ref{fig:eq_3}.
Panel (a) of the figure shows the PDF of normalized $\chi$ fluctuations, Eq.\ \eqref{eq:fluctuations}, as obtained from numerical simulations of Eq.\ \eqref{eq:eom}, for two different times, an early and a long one. At short times ($t\lesssim 1$), the PDF is well described by a Gaussian form, which is indicative of the small relevance of nonlinear effects \cite{Krug1997} due to the relatively small values of the average surface slope at those times. Indeed, the roughness remains $w(t) \lesssim 1$ during such a time regime, see Fig.\ \ref{fig:eq_1}(b). However, for long enough times, and in spite of the fact that the amplitude of the nonlinearity decreases monotonously with time in Eq.\ \eqref{eq:eom}, the PDF evolves into the characteristic TW-GOE form. Further detail on this behavior is given in the inset of Fig.\ \ref{fig:eq_3}(a), where we show the time evolution of the skewness and excess kurtosis of front fluctuations, which get close to their 1D KPZ values for times $t\gtrsim50$. The numerical cumulant values ultimately decrease for the longest times prior  to the eventual steady-state saturation of the system \cite{Rodriguez-Fernandez2020}.

Finally, and in parallel with the behavior of the one-point statistics (front PDF), the two-point statistics estimated by the front covariance also shows a nontrivial time evolution for Eq.\ \eqref{eq:eom}. This is seen in Fig.\ \ref{fig:eq_3}(b), which shows the behavior of the data collapse described in Eq.\ \eqref{eq:airy1} for different values of time. In parallel with the behavior just discussed for panel (a) of the same figure, 1D KPZ behavior, i.e.\ convergence to the Airy$_1$ covariance, is achieved in the growth regime. We can note here that Airy$_1$ covariances have been also recently found \cite{Carrasco2019} for the linear Edwards-Wilkinson (EW) equation \cite{Barabasi1995,Krug1997} in which the 1D KPZ is remarkably absent, perhaps related with the fluctuation-dissipation relation which the KPZ equation satisfies exceptionally in 1D. However, in that case the PDF is Gaussian, as the EW equation is linear, while in our present case the PDF is also of the (TW) KPZ form.

Summarizing, the scaling behavior just discussed for Eq.\ \eqref{eq:eom} is qualitatively quite similar to that found in our high-$T$ kMC simulations of the discrete model, Eq.\ \eqref{eq:energy}, including:

{\em (i)} A Washburn-like law $\langle \bar{h}(t) \rangle \sim t^{\delta}$ for the average front position, with an exponent value close to that of diffusive behavior.

{\em (ii)} A non-KPZ set of values for the scaling exponents $\alpha$, $z$, and $\beta$, in spite of the occurrence of KPZ-like nonlinear behavior at the front.

{\em (iii)} Fluctuation statistics (PDF and covariance) of the front as for the 1D KPZ universality class, in spite of {\em (ii)}. This result is to be noted as it strengthens the general conclusion (see e.g.\ Ref.\ \cite{Rodriguez-Fernandez2021} and references therein) that the precise characterization of a kinetic roughening universality class requires characterizing both traits (set of scaling exponents and fluctuation statistics) explicitly, as implied by previously known cases. Indeed, KPZ exponents do not imply KPZ statistics, as a suitable linear equation (hence, with a Gaussian PDF) has been shown to have KPZ exponents \cite{Saito2012}. Conversely, PDF and covariance of the KPZ type do not imply by themselves KPZ exponent values (as in our present case), as has already been shown for a family of nonlinear equations \cite{Nicoli2009} having non-KPZ exponents concurrent with Airy$_1$ fluctuations \cite{Nicoli2013b,Santalla2021}.

Nevertheless, differences exist between the discrete and the continuum models, the most salient ones being: {\em (i')} the precise numerical values of the scaling exponents, most notably $\beta$ and $z$, and {\em (ii')} FV scaling for the continuum model vs intrinsic anomalous scaling in the discrete model:

{\em (i')} As we noted in Sec.\ \ref{sec:eq}, Eq.\ \eqref{eq:eom} is an approximation of the full stochastic equation derived in Ref.\ \cite{Abraham2002}, which contains additional deterministic (linear and nonlinear) terms, as well as additional contributions to the noise. While incorporating some of those contributions explicitly may increase the value of the dynamic exponent $z$, hence decreasing the value of the growth exponent $\beta$, bringing both exponent values numerically closer to those obtained in our high-$T$ kMC simulations, they make simulations much more costly while not changing the main qualitative features from the point of view of scaling behavior.

{\em (ii')} In other interface growth systems in which transport is also limited by diffusion as in our present system, such as thin film production by chemical vapor or electrochemical deposition, discrete models [analogous of Eq.\ \eqref{eq:energy} here] also display intrinsic anomalous kinetic roughening \cite{Castro1998,Castro2000}, while their continuum counterparts [analogous of Eq.\ \eqref{eq:eom} here] conspicuously display FV scaling \cite{Cuerno2001,Nicoli2008,Castro2012}, in full parallel with our present case. In this context, the occurrence of anomalous scaling is expected to be a non-asymptotic feature of the discrete model, which is currently believed not to be a property of the true asymptotic scaling \cite{Lopez2005}. For example, simulations of a discrete model of diffusion-limited growth showed a crossover at extremely long times in which anomalous scaling was seen to be followed by FV scaling asymptotically \cite{Castro1998,Castro2000}. At any rate, anomalous scaling might perfectly well be found as an extremely long transient for (finite) experimental systems, as has been the case in electrodeposition of thin films \cite{Nicoli2009}.

\section{Conclusions}\label{sec:concl}

Summarizing, we have studied the spatiotemporal behavior of the fronts of liquid drops which spread out on planar substrates by means of numerical simulations. We have considered a discrete model of the system based on the Ising lattice gas, whose behavior with parameters (Hamaker constant, i.e.\ wettability, and temperature) we have addressed via extensive kinetic Monte Carlo simulations. Specifically in the context of behavior seen at high temperatures, we have additionally considered numerical simulations of a continuum stochastic equation for the front.

For wide ranges of the model parameters, we have studied classic morphological observables like the mean position of the front and its roughness. In addition, we have systematically studied two-point correlation functions in real and in Fourier space together with their time evolution, and we have assessed the statistics (probability distribution function) of the fluctuations of the front, which are unavoidable at the small physical scales associated with the process.

We can sum up the main results that we have obtained for the discrete lattice gas model:

\begin{itemize}
\item The scaling properties of the fronts of the precursor and supernatant layers are the same.
\item The $\delta\approx 0.50$ value of the exponent characterizing the mean position of the front seems to be universal for all the parameter values considered.
\item The front displays intrinsic anomalous scaling irrespective of parameter values, in such a way that the roughness exponents quantifying front fluctuations at large ($\alpha$) and small ($\alpha_{\rm loc}$) length scales are different.
\item The critical exponent values $\beta$, $\alpha$, and $z$ depend more strongly on temperature than on the Hamaker constant.
\item The values of the critical exponents show a transition from a low-temperature to a high-temperature regime.
\item For the lowest temperatures we have studied, we obtain exponent values which are close to those previously reported for the same model \cite{Abraham2002}: $\alpha\simeq 0.6$, $\alpha_{\rm loc}\simeq 0.38$, $z\simeq3.3$, and $\beta\simeq0.18$.
\item The exponent values change rapidly with $T$ and become $T$-independent for $T\gtrsim 1$ at $\alpha\simeq 0.90$, $\alpha_\mathrm{loc} \simeq 0.45$, $z \simeq 3.3$, and $\beta\simeq 0.26$.
\item In spite of the exponent values and of the intrinsic anomalous scaling behavior, the statistics of front fluctuations (height PDF and covariance) agree with those characteristic of the one-dimensional KPZ universality class for $T\gtrsim T_*=3/4$.
\end{itemize}

Our results on critical exponent values agree with (and generalize) those reported by different groups of authors for the same model \cite{Lukkarinen1995,Abraham2002,Luo2019}, while differing from other \cite{Harel2018,Harel2021} which feature non-monotonic behavior of the exponents with system parameters. We find it reassuring that the scaling behavior we obtain (in particular, the $R(t)\sim t^{1/2}$ law) coincides with results from alternative modeling and simulation approaches \cite{DeConinck2008,Bonn2009,Popescu2012}, and in particular with the continuum evolution equation for the front, Eq.\ \eqref{eq:eom}. With respect to this, let us summarize the main aspects:

\begin{itemize}
\item Equation \eqref{eq:eom} features exponent values not unlike those obtained in the high-temperature regime of our discrete model, $\delta\simeq 0.58$, $\alpha\simeq 1$, $z\simeq2.65$, and $\beta\simeq0.37$, with deviations being larger for $\delta$, and especially $z$ and $\beta$.

\item These exponents are close to, but differ from, those of the linearized version of Eq.\ \eqref{eq:eom} in which $\lambda=0$, which are $\alpha=1$ and $z=3$. For such a linear equation, the front does not move on average, i.e., $\langle \bar{h}(t) \rangle=0$ \cite{Krug1997}.

\item The scaling behavior displayed by Eq.\ \eqref{eq:eom} is not anomalous; in particular,
$\alpha_{\rm loc} = \alpha\simeq 1$.

\item Although the scaling exponents predicted by Eq.\ \eqref{eq:eom} differ from those of the 1D KPZ equation, the front statistics (PDF and covariance) are those of the 1D KPZ universality class.

\end{itemize}

In general, we believe that the comparison between the discrete and the continuous models that we have simulated underscores the occurrence of universal behavior in the spreading of thin fluid films, which is particularly clear-cut at high temperatures. Surprisingly, the front fluctuations in such a regime displays properties that make it qualify as another instance of 1D KPZ behavior (although not of KPZ exponents!), as is currently being found in the dynamics of many low-dimensional, strongly-correlated, nonequilibrium systems \cite{Takeuchi2018}. It would be extremely interesting to assess if this conclusion from our ``microscopic'' simulations can be confirmed by simulations of a different nature (e.g., molecular dynamics or lattice-Boltzmann approaches), and/or in experiments on spreading of precursor films.

\begin{acknowledgments}

This work was partially supported by Ministerio de Econom\'ia, Industria y Competitividad (MINECO, Spain), Agencia Estatal de Investigaci\'on (AEI, Spain), and Fondo Europeo de Desarrollo Regional (FEDER, EU) through Grants PID2020-112936GB-I00 and PGC2018-094763-B-I00, by the Junta de Extremadura (Spain) and Fondo Europeo de Desarrollo Regional (FEDER, EU) through Grant No.\ GRU18079 and IB20079, and by Comunidad de Madrid (Spain) under the Multiannual Agreement with UC3M in the line of Excellence of University Professors (EPUC3M23), in the context of the 5th.\ Regional Programme of Research and Technological Innovation (PRICIT). J.\ M.\ was supported by Programa Propio de Iniciación a la Investigación de la Universidad de Extremadura through Scolarship No.\ 1362. P.\ R.-L.\ was supported by "AYUDA PUENTE 2021, URJC".

Our kMC simulations have been performed in the computing facilities of the Instituto de Computaci\'{o}n Cient\'{\i}fica Avanzada de Extremadura (ICCAEx).

\end{acknowledgments}

\bibliography{ThinFilm.bib}

\onecolumngrid
\appendix
\section{Simulation parameters and details} \label{details}

In the Kawasaki local dynamics two adjacent cells are selected and their values are exchanged following the Metropolis acceptance criterion, namely \cite{Newman1999}
\begin{equation}
    A(\mu \rightarrow \nu)=\left\{\begin{array}{cc}{e^{-\beta \Delta E}} & {\Delta E>0}\,, \\ {1} & {\Delta E \leq 0}\,,\end{array}\right.
\end{equation}
where $A(\mu \rightarrow \nu)$ is the acceptance rate for the $\mu \rightarrow \nu$ transition, $\beta=1/k_{B}T$ and $\Delta E=E_{\nu}-E_{\mu}$ is the energy difference between the final and initial states. When the temperature is very low, the system may eventually get pinned in a given state due to a very low acceptance rate. Moreover, it is possible to select two cells whose occupations are the same, in which case the system remains the same upon exchange. To overcome these problems we use a continuous time algorithm in which we track all the non-trivial exchanges that may take place. The starting point for the algorithm is to select one of these exchanges, proportionally to their acceptance ratio, and to carry out the exchange. The simulation time is then updated by adding the time interval \cite{Newman1999}
\begin{equation}
    \Delta t=\frac{-1}{\log P(\mu \rightarrow \mu)}.
\end{equation}
The probability of permanence in the initial state $P(\mu \rightarrow \mu)$ can be calculated as:
\begin{equation}
    P(\mu \rightarrow \mu)=\frac{1}{N_P}\left[\#\mathrm{Trivial \hspace{2mm}Exchanges} +\sum_{\nu\ne\mu}(1-A(\mu \rightarrow \nu))\right],
    \label{eq:prob_remain}
\end{equation}
where $N_P$ is the number of adjacent pairs in our system. To compute $P(\mu \rightarrow \mu)$ we must take into account all the trivial exchanges (between empty or full cells) that are accepted always but leave invariant the system, as well as all the non-trivial exchanges that can be rejected when proposed. Therefore, Eq.~\eqref{eq:prob_remain} may then be reduced to
\begin{equation}
    P(\mu \rightarrow \mu)=\frac{1}{N_P}\left[N_P -\sum_{\nu\ne\mu}A(\mu \rightarrow \nu)\right]\,.
    \label{eq:prob_remain2}
\end{equation}
In our simulations we do not fix the final time, but the total number of exchanges that will take place. As each run has a different seed, times between runs are not the same.

In all runs $J$ was fixed to $1$, so that we only modified the value of the Hamaker constant $A$ and the temperature $T$. As we are interested in setups where both terms in the energy [see Eq.~\eqref{eq:energy}] are dominant, we selected a wide range of values of the Hamaker constant and of temperature. Other than fixing $k_B=1$, we keep the units of $A$ and $T$ arbitrary, as we are only interested in the values of the ratios $J/{k_{B}T}$ and $A/{k_{B}T}$, that control the acceptance rates. We fixed $L_x = 1000$ in all runs, which was long enough to avoid the film to reach the edge of the system. On the other hand, we choose $L_y=256$ in most of the simulations. In Table \ref{tab:det} we report all the simulation conditions that we have considered.

Additionally, Tables \ref{tab:delta} and \ref{tab:beta} contain the results obtained for the $\delta$ and $\beta$ exponents defined in the main text, as measured from $\langle \bar{h}(t) \rangle$
and $w(t)$, respectively. In these tables, given an exponent $\nu$, we use $\nu_1$ to denote its value for the front of the precursor film ($Z=1$) and $\nu_2$ to denote its value for the front of the supernatant layer ($Z=2$). Tables \ref{tab:a8} and \ref{tab:a9} collect the results for the dynamic and roughness exponents obtained for the front of the precursor layer by analyzing the height-difference correlation function $C_2(r,t)$ as detailed in section \ref{sec:hdcf} of the main text. Finally, Table \ref{tab:ap} collects the values of $2\alpha'=2(\alpha-\alpha_{\rm loc})$ as obtained from data collapses of the height-difference correlation function analogous of that shown in Fig.\ \ref{fig:c2}. Note $\alpha'\neq0$, implying intrinsic anomalous scaling for all parameter values.

\begin{table}[!hbp]
\begin{ruledtabular}
\begin{tabular}{ c | c | c | c | c | c }
$L_x$ & $L_y$ & $T$ & $A$ & $N_E$ & Number of runs \\ \hline\hline
 &  & 10 & 10 & $1.5 \times 10^8$ & 100 \\ \cline{3 -6}
  &  & 10 & 5 & $1.5 \times 10^8$ & 100 \\ \cline{3 -6}
1000 & 256 & 10 & 1 & $1.0 \times 10^8$ & 100 \\ \cline{ 3- 6}
  &  & 10 & 0.1 & $1.0 \times 10^8$ & 100 \\ \cline{3 -6}
    &  & 10 & 0.01 & $1.0 \times 10^8$ & 100 \\ \hline

 &  & 3 & 10 & $1.0 \times 10^8$ & 100 \\ \cline{3 -6}
  &  & 3 & 5 & $1.0 \times 10^8$ & 100 \\ \cline{3 -6}
1000 & 256 & 3 & 1 & $1.0 \times 10^8$ & 100 \\ \cline{ 3- 6}
  &  & 3 & 0.1 & $1.0 \times 10^8$ & 100 \\ \cline{3 -6}
    &  & 3 & 0.01 & $1.0 \times 10^8$ & 100 \\ \hline

 &  & 1 & 10 & $2.0 \times 10^8$ & 100 \\ \cline{3 -6}
  &  & 1 & 5 & $2.0 \times 10^8$ & 100 \\ \cline{3 -6}
1000 & 256 & 1 & 1 & $2.0 \times 10^8$ & 1000 \\ \cline{ 3- 6}
  &  & 1 & 0.1 & $2.0 \times 10^8$ & 100 \\ \cline{3 -6}
    &  & 1 & 0.01 & $2.0 \times 10^8$ & 100 \\ \hline

 &  & 3/4 & 10 & $4.0 \times 10^8$ & 100 \\ \cline{3 -6}
  &  & 3/4 & 5 & $4.0 \times 10^8$ & 100 \\ \cline{3 -6}
1000 & 256 & 3/4 & 1 & $4.0 \times 10^8$ & 100 \\ \cline{ 3- 6}
  &  & 3/4 & 0.1 & $4.0 \times 10^8$ & 100 \\ \cline{3 -6}
    &  & 3/4 & 0.01 & $4.0 \times 10^8$ & 100 \\ \hline

 &  & 1/2 & 10 & $7.5 \times 10^8$ & 100 \\ \cline{3 -6}
  &  & 1/2 & 5 & $7.5 \times 10^8$ & 100 \\ \cline{3 -6}
1000 & 256 & 1/2 & 1 & $7.5 \times 10^8$ & 100 \\ \cline{ 3- 6}
  &  & 1/2 & 0.1 & $7.5 \times 10^8$ & 100 \\ \cline{3 -6}
    &  & 1/2 & 0.01 & $7.5 \times 10^8$ & 100 \\ \hline

 &  & 1/3 & 1 & $1.25 \times 10^{10}$ & 100 \\ \cline{3 -6}
1000 & 256 & 1/3 & 0.1 & $1.25 \times 10^{10}$ & 100 \\ \cline{ 3- 6}
  &  & 1/3 & 0.01 & $1.25 \times 10^{10}$ & 100 \\ \hline

1000 & 64 & 1/3 & 10 & $5.0 \times 10^9$ & 100 \\ \hline
1000 & 64 & 1/3 & 5 & $5.0 \times 10^9$ & 100 \\ \hline
1000 & 128 & 1 & 1 & $1.0 \times 10^8$ & 250 \\ \hline
1000 & 512 & 1 & 1 & $4.0 \times 10^8$ & 250 \\
\end{tabular}
\end{ruledtabular}
\caption{Parameters used for the runs reported herein. $N_E$ is the total number of the exchanges performed, and the last column shows the number of runs launched in each case.}
\label{tab:det}
\end{table}

\begin{table}[!hbp]
\begin{ruledtabular}
\begin{tabular}{|c |c| c|c |c|c| c|c| c|c| c|}
\multirow{2}{*}{\diagbox[width=2.5em]{$A$}{$T$}} & \multicolumn{2}{c |}{10} & \multicolumn{2}{c|}{3}  & \multicolumn{2}{c|}{1}  & \multicolumn{2}{c|}{3/4}  & \multicolumn{2}{c|}{1/2} \\ \cline{2-11}

 & $\delta_1$ & $\delta_2$ & $\delta_1$ & $\delta_2$ & $\delta_1$ & $\delta_2$ & $\delta_1$ & $\delta_2$ & $\delta_1$ & $\delta_2$ \\ \hline
 \hline
         10 & 0.4804(7) & 0.471(1) & 0.4911(4) & 0.4721(8) & 0.5091(4) & 0.4887(9) & 0.5165(4) & 0.4945(9) & 0.5503(4) & 0.491(1) \\ \hline
         5 & 0.4781(8) & 0.472(1) & 0.485(6) & 0.4719(9) & 0.5079(5) & 0.4892(9) & 0.5169(3) & 0.4952(9) & 0.5502(4) & 0.490(1) \\ \hline
         1 & 0.4751(9) & 0.4742(9) & 0.4799(5) & 0.4771(5) & 0.489(1) & 0.493(1) & 0.4891(8) & 0.5061(7) & 0.4985(4) & 0.5181(4) \\ \hline
         0.1 & 0.474(1) & 0.474(9) & 0.4798(6) & 0.4792(6) & 0.4909(8) & 0.4929(8) & 0.4897(6) & 0.4920(6) & 0.5103(5) & 0.5113(5) \\  \hline
         0.01 & 0.4754(9) & 0.4753(9) & 0.4766(8) & 0.4768(8) & 0.4889(9) & 0.4890(9) & 0.4931(6) & 0.4933(6) & 0.5131(7) & 0.5132(7) \\ 
\end{tabular}
\end{ruledtabular}
\caption{Values of the exponents $\delta_1$ and $\delta_2$, for the precursor and supernatant layers, respectively, for all the conditions studied.}
\label{tab:delta}
\end{table}
\bigskip

\begin{table}[!t]
\begin{ruledtabular}
\begin{tabular}{|c |c| c|c |c|c| c|c| c|c| c|}
\multirow{2}{*}{\diagbox[width=2.5em]{$A$}{$T$}} & \multicolumn{2}{c |}{10} & \multicolumn{2}{c|}{3}  & \multicolumn{2}{c|}{1}  & \multicolumn{2}{c|}{3/4}  & \multicolumn{2}{c|}{1/2} \\ \cline{2-11}
 & $2\beta_1$ & $2\beta_2$ & $2\beta_1$ & $2\beta_2$ & $2\beta_1$ & $2\beta_2$ & $2\beta_1$ & $2\beta_2$ & $2\beta_1$ & $2\beta_2$ \\ \hline
 \hline
 10 & 0.539(4) & 0.538(3) & 0.536(3) & 0.541(3) & 0.516(7) & 0.530(3) & 0.489(8) & 0.489(4) & 0.29(1) & 0.318(9) \\ \hline
 5 & 0.537(4) & 0.537(3) & 0.533(3) & 0.539(3) & 0.517(7) & 0.526(4) & 0.483(6) & 0.479(4) & 0.28(2) & 0.314(8) \\ \hline
 1 & 0.536(4) & 0.533(4) & 0.538(3) & 0.538(3) & 0.544(8) & 0.537(8) & 0.536(9) & 0.503(8) & 0.26(1) & 0.29(1) \\ \hline
 0.1 & 0.543(4) & 0.542(4) & 0.538(3) & 0.536(3) & 0.475(9) & 0.476(9) & 0.343(9) & 0.347(9) & 0.30(2) & 0.30(2) \\  \hline
 0.01 & 0.537(4) & 0.539(4) & 0.538(4) & 0.538(4) & 0.497(9) & 0.494(9) & 0.34(1) & 0.34(1) & 0.33(3) & 0.34(3) \\ 
\end{tabular}
\end{ruledtabular}
\caption{Just as in Table \ref{tab:delta} for the growth exponents $2\beta_1$ and $2\beta_2$.}
\label{tab:beta}
\end{table}

\begin{table}[!t]
\begin{ruledtabular}
\begin{tabular}{|c |c| c|c |c|c| c|c| c|c| c|}
\multirow{2}{*}{\diagbox[width=2.5em]{$A$}{$T$}} & \multicolumn{2}{c |}{10} & \multicolumn{2}{c|}{3}  & \multicolumn{2}{c|}{1}  & \multicolumn{2}{c|}{3/4}  & \multicolumn{2}{c|}{1/2} \\ \cline{2-11}
 & $1/z$ & $2\alpha$ & $1/z$ & $2\alpha$ & $1/z$ & $2\alpha$ & $1/z$ & $2\alpha$ & $1/z$ & $2\alpha$ \\ \hline
 \hline
         10 & 0.299(5) & 1.83(2) & 0.309(5) & 1.77(2) & 0.286(3) & 1.78(2) & 0.251(6) & 1.92(3) & 0.202(9) & 1.33(5) \\ \hline
         5 & 0.299(5) & 1.83(2) & 0.304(6) & 1.79(3) & 0.289(3) & 1.76(2) & 0.224(6) & 1.95(4) & 0.203(8) & 1.35(4) \\ \hline
         1 & 0.299(6) & 1.83(3) & 0.304(3) & 1.81(2) & 0.303(4) & 1.78(2) & 0.259(6) & 2.06(5) & 0.27(1) & 1.15(5) \\ \hline
         0.1 & 0.309(5) & 1.80(2) & 0.304(3) & 1.81(1) & 0.257(4) & 1.89(3) & 0.236(6) & 1.43(3) & 0.28(2) & 1.07(5) \\  \hline
         0.01 & 0.301(5) & 1.82(3) & 0.307(5) & 1.79(3) & 0.260(5) & 1.88(3) & 0.245(9) & 1.40(5) & 0.25(2) & 1.23(6) \\ 
\end{tabular}
\end{ruledtabular}
\caption{Values of the exponents $1/z$ and $2\alpha$ for the precursor layer, calculated with $a=0.8$.} \label{tab:a8}
\end{table}
\bigskip

\begin{table}[!t]
\begin{ruledtabular}
\begin{tabular}{|c |c| c|c |c|c| c|c| c|c| c|}
\multirow{2}{*}{\diagbox[width=2.5em]{$A$}{$T$}} & \multicolumn{2}{c |}{10} & \multicolumn{2}{c|}{3}  & \multicolumn{2}{c|}{1}  & \multicolumn{2}{c|}{3/4}  & \multicolumn{2}{c|}{1/2} \\ \cline{2-11}
 & $1/z$ & $2\alpha$ & $1/z$ & $2\alpha$ & $1/z$ & $2\alpha$ & $1/z$ & $2\alpha$ & $1/z$ & $2\alpha$ \\ \hline
 \hline
 10 & 0.296(6) & 1.85(4) & 0.305(7) & 1.79(3) & 0.285(4) & 1.79(3) & 0.251(8) & 1.91(6) & 0.22(3) & 1.2(1) \\ \hline
 5 & 0.294(7) & 1.85(4) & 0.301(6) & 1.80(3) & 0.288(5) & 1.76(3) & 0.247(7) & 1.93(5) & 0.21(1) & 1.30(7) \\ \hline
 1 & 0.295(6) & 1.86(3) & 0.299(3) & 1.84(2) & 0.307(5) & 1.75(3) & 0.261(9) & 2.04(7) & 0.30(3) & 1.0(1) \\ \hline
 0.1 & 0.306(8) & 1.81(4) & 0.301(4) & 1.82(3) & 0.258(5) & 1.87(4) & 0.245(9) & 1.36(6) & 0.3(1) & 0.7(3) \\  \hline
 0.01 & 0.296(7) & 1.85(4) & 0.304(7) & 1.80(4) & 0.263(7) & 1.85(5) & 0.25(2) & 1.33(8) & 0.18(4) & 1.5(3) \\ 
\end{tabular}
\end{ruledtabular}
\caption{The same as in Table \ref{tab:a8}, but for $a=0.9$.}
\label{tab:a9}
\end{table}

\begin{center}
\begin{table}[!t]
\begin{ruledtabular}
\begin{tabular}{|c|c|c|c|c|c|}
\diagbox[width=2.5em]{$A$}{$T$} & 10 & 3 & 1 & 3/4 & 1/2 \\ \hline
 \hline
        10 & 0.89(2)&0.86(3) &0.87(2) &1.00(3) &0.44(5) \\ \hline
        5 & 0.90(2)& 0.86(2)& 0.85(2)& 1.03(4)& 0.47(3)\\ \hline
        1 & 0.89(3)& 0.88(1)& 0.87(2)& 1.13(4)& 0.36(5)\\ \hline
        0.1 & 0.87(2)& 0.88(1)& 1.01(3)& 0.61(3)& 0.37(5)\\  \hline
        0.01 & 0.89(3)& 0.87(3)& 0.99(3)&0.60(5) &0.53(6) \\ 
\end{tabular}
\end{ruledtabular}
\caption{Values of the exponents $2\alpha'$ for the precursor layer, calculated with $a=0.8$.}
\label{tab:ap}
\end{table}
\end{center}

\end{document}